\documentclass[reprint, amsmath,amssymb, prl, floatfix, longbibliography]{revtex4-1} 
\usepackage{graphicx}% Include figure files
\usepackage{bm}% bold math
\usepackage{color}
\usepackage{longtable}
   \makeatletter
     \renewcommand\@make@capt@title[2]{%
      \@ifx@empty\float@link{\@firstofone}{\expandafter\href\expandafter{\float@link}}%
       {\textbf{#1}}\@caption@fignum@sep#2\quad}%
     \makeatother

\begin{document}
%\linenumbers
\makeatletter 
\renewcommand{\fnum@figure}{\textbf{Figure~\thefigure}}
\makeatother

% ****** Start of file apssamp.tex ******
%
%   This file is part of the APS files in the REVTeX 4.1 distribution.
%   Version 4.1r of REVTeX, August 2010
%
%   Copyright (c) 2009, 2010 The American Physical Society.
%
%   See the REVTeX 4 README file for restrictions and more information.
%
% TeX'ing this file requires that you have AMS-LaTeX 2.0 installed
% as well as the rest of the prerequisites for REVTeX 4.1
%
% See the REVTeX 4 README file
% It also requires running BibTeX. The commands are as follows:
%
%  1)  latex apssamp.tex
%  2)  bibtex apssamp
%  3)  latex apssamp.tex
%  4)  latex apssamp.tex
%
%\documentclass[%
% reprint,
%superscriptaddress,
%groupedaddress,
%unsortedaddress,
%runinaddress,
%frontmatterverbose,
%preprint,
%showpacs,preprintnumbers,
%nofootinbib,
%nobibnotes,
%bibnotes,
% amsmath,amssymb,
% aps,
%prl
%pra,
%prb,
%rmp,
%prstab,
%prstper,
%floatfix,
%]{revtex4-1}

%\usepackage{graphicx}% Include figure files
%\usepackage{dcolumn}% Align table columns on decimal point
%\usepackage{bm}% bold math
%\usepackage{amsmath}
%\usepackage{hyperref}% add hypertext capabilities
%\usepackage[mathlines]{lineno}% Enable numbering of text and display math
%\linenumbers\relax % Commence numbering lines

%\usepackage[showframe,%Uncomment any one of the following lines to test
%%scale=0.7, marginratio={1:1, 2:3}, ignoreall,% default settings
%%text={7in,10in},centering,
%%margin=1.5in,
%%total={6.5in,8.75in}, top=1.2in, left=0.9in, includefoot,
%%height=10in,a5paper,hmargin={3cm,0.8in},
%]{geometry}

%\begin{document}

\preprint{APS/123-QED}

%\title{Triplon band splitting and topologically protected edge states in the dimerized antiferromagnet Ba$_2$CuSi$_2$O$_6$Cl$_2$} % Force line breaks with \\
\title{Triplon band splitting and topologically protected edge states in the dimerized antiferromagnet} % Force line breaks with \\
\author{Kazuhiro Nawa$^1$}
\email{knawa@tohoku.ac.jp}
\author{Kimihiko Tanaka$^2$}
\author{Nobuyuki Kurita$^2$}
\author{Taku J Sato$^1$}
\author{Haruki Sugiyama$^3$}
\author{Hidehiro Uekusa$^3$}
\author{Seiko Ohira-Kawamura$^4$}
\author{Kenji Nakajima$^4$}
\author{Hidekazu Tanaka$^2$}
\email{tanaka@lee.phys.titech.ac.jp}

\affiliation{
$^1$Institute of Multidisciplinary Research for Advanced Materials, Tohoku University, 2-1-1 Katahira, Sendai 980-8577, Japan \\
$^2$Department of Physics, Tokyo Institute of Technology, Meguro-ku, Tokyo 152-8551, Japan\\
$^3$Department of Chemistry, Tokyo Institute of Technology, Meguro-ku, Tokyo 152-8551, Japan\\
$^4$Materials and Life Science Division, J-PARC Center, Tokai, Ibaraki 319-1195, Japan
}

\date{\today}% It is always \today, today,

\begin{abstract}
The search for topological insulators has been actively promoted in the field of condensed matter physics for further development in energy-efficient information transmission and processing.
In this context, recent studies have revealed that not only electrons but also bosonic particles such as magnons can construct edge states carrying nontrivial topological invariants.
Here we demonstrate topological triplon bands in the spin-1/2 two-dimensional dimerized quantum antiferromagnet Ba$_2$CuSi$_2$O$_6$Cl$_2$, which is closely related to a pseudo-one-dimensional variant of the Su-Schrieffer-Heeger (SSH) model,
through inelastic neutron scattering experiments.
The excitation spectrum exhibits two triplon bands and a clear band gap between them due to a small alternation in interdimer exchange interactions along the $a$-direction, 
which is consistent with the crystal structure. The presence of topologically protected edge states is indicated by a bipartite nature of the lattice.
\end{abstract}

\maketitle

%\section{\label{Introduction}Introduction}
The discoveries of quantum Hall effects~\cite{Klitzing} and topological insulators~\cite{Hsieh} have shed light on gapless edge states that exist between phases with different topological characters~\cite{Review, Review2}.
%triggering an intense search for topological insulators that could be useful for future energy-efficient information transmission and processing.
In addition, concepts of edge states have been extended to other systems, such as ultracold atom systems in optical lattices~\cite{coldatom, coldatom2, coldatom3},
and even bosonic counterparts such as photonic crystals~\cite{photon, photon2}, phonons~\cite{phonon}, and magnons~\cite{magnon, MagnonHall2, magnon1, magnon2, magnon3} in solids.
In electron systems, the topological characters are classified by the total topological invariant of the occupied bands, which is associated with quantized conductance~\cite{TKNN, Review, Review2}.
On the other hand, for insulators, where ground and excited states are described by bosonic particles such as magnons, 
thermodynamic conductance is dominated by the topological characters of thermally excited bands~\cite{MagnonHall, MagnonHall2, MagnonHall3}.
Thus, it is necessary to reveal the detailed dispersion relations of excited bands to explore and design insulators with bosonic topological bands.
 
From these viewpoints, a dimerized magnet, which has well-defined bosonic excitations called triplons,
is a good starting point for realizing bosonic topological bands~\cite{Rice, Giamarchi,Zapf}.
%Because of dominant antiferromagnetic intradimer interactions,
%triplon bands are formed with a finite energy gap from the ground state, which is well described as a product of singlets. 
%On the other hand, interdimer exchange interactions induce the hopping (from transverse terms) and repulsion (from longitudinal terms) of triplons~\cite{Rice, Giamarchi,Zapf}.
One of the advantages of studying a dimerized magnet is that triplon bands can be easily deformed by applying a magnetic field or hydrostatic pressure.
If the deformation is so large that a triplet excitation energy is tuned to zero, quantum phase transitions occur~\cite{Giamarchi,Zapf, Matsumoto1, Matsumoto2},
leading to Bose-Einstein condensation~\cite{O_mag,Nikuni,Oosawa,Rueegg,Sasago,Jaime,Sebastian} or a Wigner crystal of triplons~\cite{Kageyama, Kageyama2, Kodama, Tanaka} induced by a magnetic field.
%For instance, with an increasing magnetic field, an $S^z\,{=}\,{+}1$ branch of triplons undergoes Bose-Einstein condensation (BEC) if the kinetic energy of triplons is more effective than repulsive interactions
%~\cite{O_mag,Nikuni,Oosawa,Rueegg,Sasago,Jaime,Sebastian},
%while a Wigner crystal of localized triplons is realized in the opposite case~\cite{Kageyama, Kageyama2, Kodama, Tanaka}.
In addition, recent theoretical works have revealed that triplon bands in a certain dimerized magnet can be regarded as topologically insulating~\cite{edgestate1, edgestate2, edgestate3}.
For instance, the $S^z\,{=}\,{+}1$, 0, and $-1$ branches of triplons can be both topologically trivial and non-trivial by controlling the magnitude of a magnetic field in SrCu$_2$(BO$_3$)
owing to interdimer Dzyaloshinskii-Moriya interactions which yield complex hopping amplitudes~\cite{edgestate1, edgestate5}.
This is supported by detailed calculation on a winding number and edge states by using exchange parameters determined from inelastic neutron scattering experiments with a very high accuracy~\cite{edgestate4}. 

%Quite interestingly, recent theoretical works have revealed that a certain dimerized magnet can be regarded as topologically insulating in terms of triplons~\cite{edgestate1, edgestate2, edgestate3}.
%For instance, the $S^z\,{=}\,{+}1$, 0, and $-1$ branches of triplons should become topologically nontrivial under a certain magnetic field in SrCu$_2$(BO$_3$)
%owing to interdimer Dzyaloshinskii-Moriya interactions which yield complex hopping amplitudes~\cite{edgestate1}.
%This is supported by detailed calculation on a winding number and edge states by using exchange parameters determined from inelastic neutron scattering experiments with a very high accuracy~\cite{edgestate4}.  
%The transition between topologically trivial and non-trivial phases can be tuned by controlling the magnitude or direction of a magnetic field~\cite{edgestate1, edgestate5}. {At the critical magnetic field between the two phases, the formation of a spin-1 Dirac cone is expected~\cite{edgestate1}.

\begin{figure}[t]
\includegraphics[width=8cm, clip]{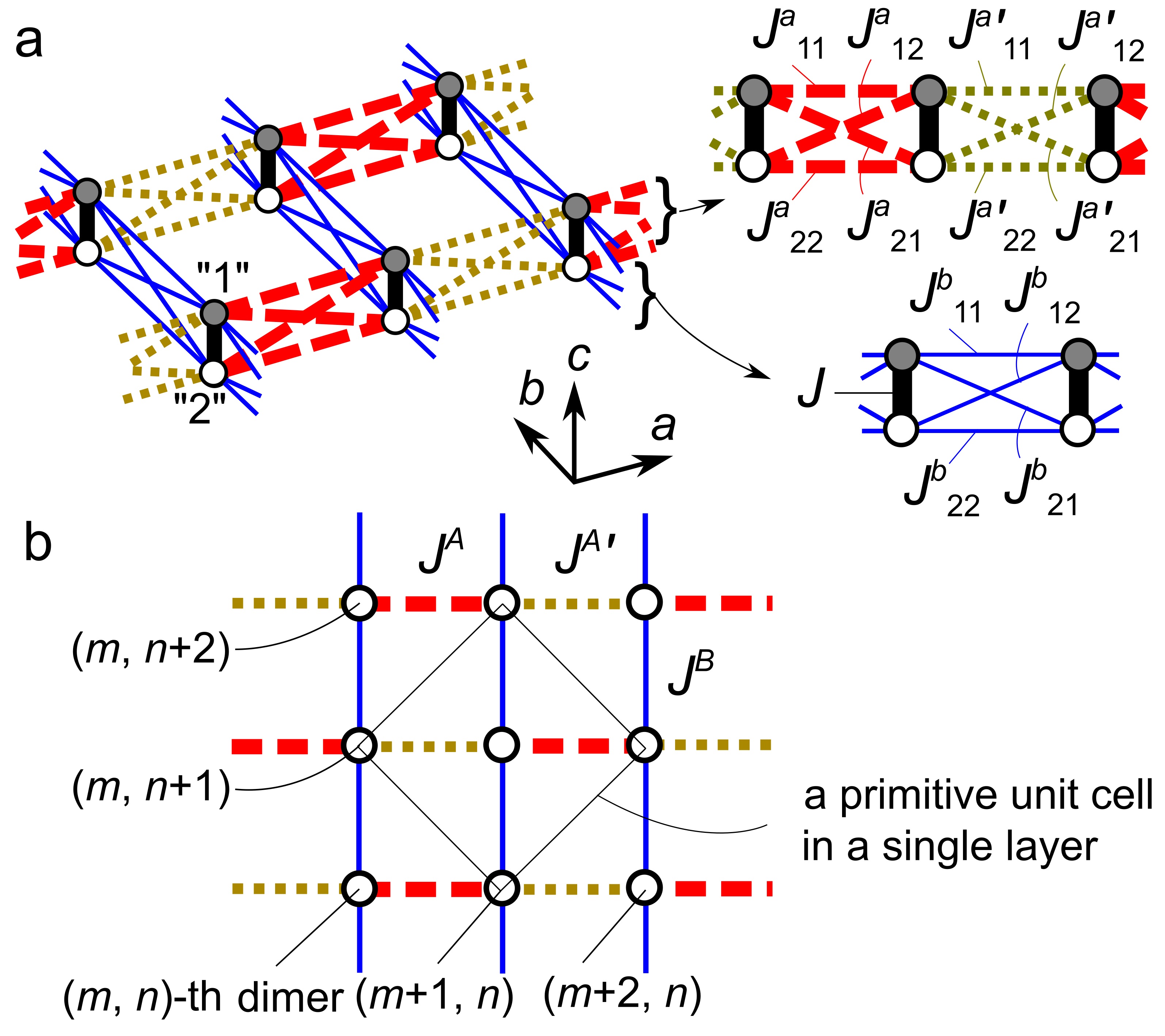} %300dpi
\caption{\label{model} {\bf Intradimer and interdimer couplings in Ba$_2$CuSi$_2$O$_6$Cl$_2$.} {\bf a.} Possible interactions between Cu atoms.
Shaded and open circles indicate two Cu atoms forming antiferromagnetic dimers.
{\bf b.} Effective couplings between dimers expected from the crystal symmetry $Cmc2_1$.
Dashed, dotted, and solid lines represent hopping amplitudes denoted by $J^{A}$, $J^{A\prime}$, and $J^B$, respectively (see eq.~\eqref{definitionofJ} for definition).}
\end{figure}

Ba$_2$CuSi$_2$O$_6$Cl$_2$ crystallizes in a layered structure with each layer composed of antiferromagnetically coupled dimers~\cite{Okada};
the structure is closely related to that of Ba$_2$CoSi$_2$O$_6$Cl$_2$~\cite{Tanaka}.
Figure~\ref{model}a illustrates the 2D exchange network of Ba$_2$CuSi$_2$O$_6$Cl$_2$.
It is slightly different from that reported previously (with the space group $Cmce$)~\cite{Okada}
and is based on the new structure with the space group $Cmc2_1$, which allows an alternation in interdimer exchanges along the $a$-axis, as we will discuss later.
A pair of nearest-neighbor Cu atoms form antiferromagnetic dimers almost parallel to the $c$-axis via exchange couplings $J$.
These dimers are coupled via interdimer exchange couplings $J_{ij}^{\alpha}$ and $J_{ij}^{\prime\alpha}$ with $i, j$ =~1, 2 and $\alpha =~a, b$, forming a 2D exchange network in the $ab$ plane.
In fact, the magnetic properties of Ba$_2$CuSi$_2$O$_6$Cl$_2$ are well characterized by a spin-1/2 quasi-2D dimer system~\cite{Okada}.
%The magnetization curve is excellently reproduced using the exact diagonalization calculation based on the 2D coupled dimer model, indicating strongly two-dimensional characters in the exchange network.
Under the assumption of $J_p \equiv J^\alpha_{11} = J^{\alpha\prime}_{11} = J^\alpha_{22} = J^{\alpha\prime}_{22}$ and $J_d \equiv J^\alpha_{12} = J^{\alpha\prime}_{12} = J^\alpha_{21} = J^{\alpha\prime}_{21}$ ($\alpha = a, b$),
the exchange constants are estimated as $J$ =~2.42~meV, $J_p$ =~0.03~meV, $J_d$ =~0.34~meV from the magnetization curve and density functional theory calculations~\cite{Okada}.
%In addition, the magnetic anisotropy should be very small since magnetic susceptibilities and entire magnetization curves for two different field orientations coincide almost perfectly with each other after being normalized by the $g$-factors.

\begin{figure*}[t]
\includegraphics[width=16cm, clip]{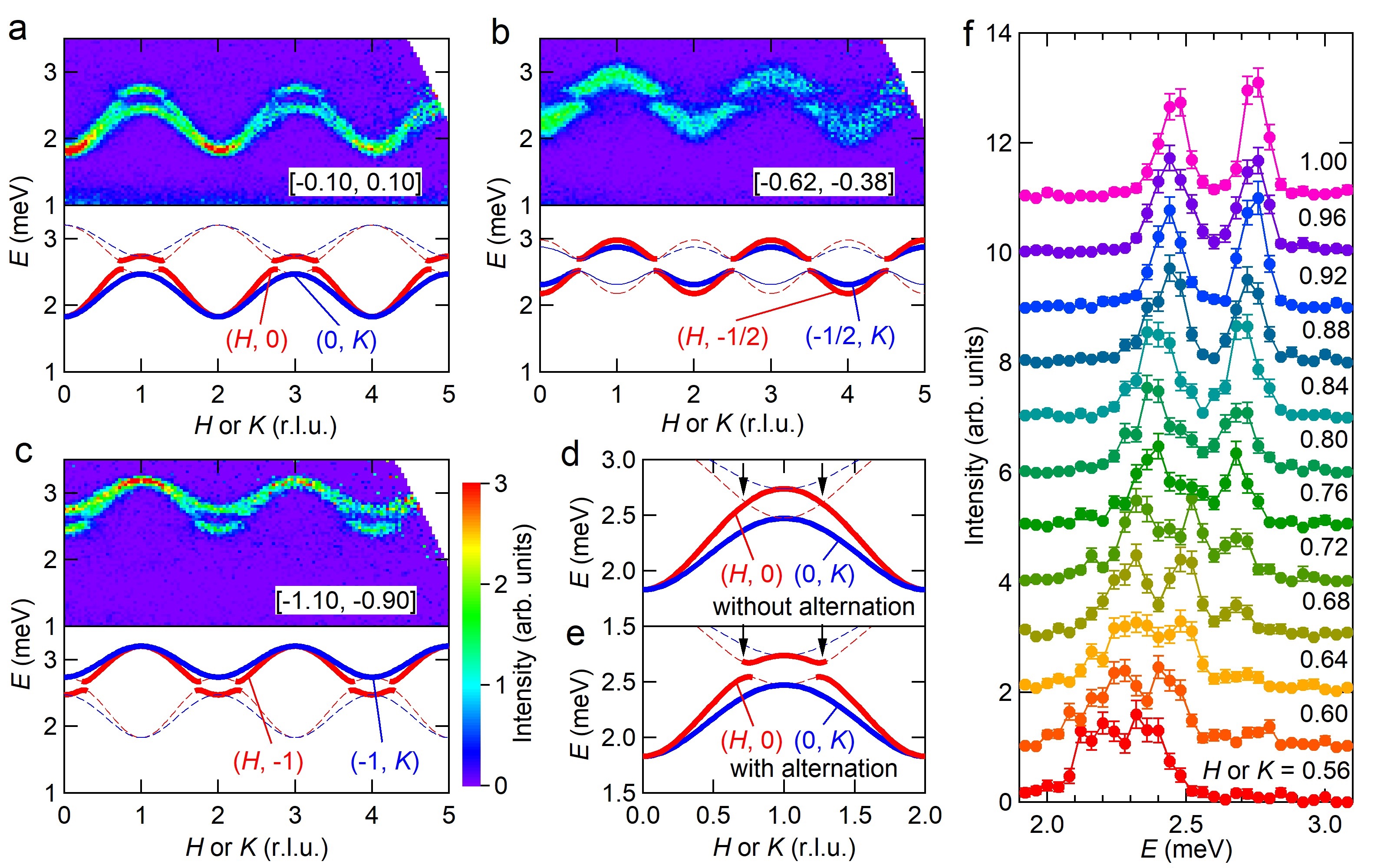} %300dpi
\caption{\label{dispersion1}
{\bf Excitation spectra along H or K directions.}
{\bf a.} Energy-momentum maps of the scattering intensities along the ($H$, 0) direction (shown above) compared with the dispersion relations of eq.~\eqref{diseq} (below).
Intensity is integrated for the $K$ range shown in the figure ($\Delta K$ =~$\pm$0.10) and the entire observed $L$ range.
The branch along the (0, $K$) direction is observed as well as that along the ($H$, 0) direction because of crystallographic domains.
{\bf b, c.} The same maps along ($H$, -1/2) and ($H$, -1) directions and the corresponding dispersion relations.
The same color scale is used for all the maps.
The presented data are collected at 0.3~K with an $E_i$ of 5.92 meV. 
Thick solid and thin dashed curves indicate triplon bands with a large and small structure factor, respectively.
{\bf d, e} Expected triplon bands without and with an alternation in interdimer interactions along the ($H$, 0) and (0, $K$) directions.
Arrows indicate a crossing point of two triplon bands when the alternation is absent.
{\bf f.} Intensity plotted as a function of $E$ at selected reciprocal space points ($H$, 0) and (0, $K$) in Fig.~\ref{dispersion1}a.}
\end{figure*}

The main finding of this study is the gap between two triplon bands, as shown in Fig.~\ref{dispersion1}.
As we discuss later, this contradicts the previously reported crystal structure which does not yield
two triplon bands or the gap between them under the crystal symmetry of $Cmce$.
Thus, we reinvestigated the crystal structure of Ba$_2$CuSi$_2$O$_6$Cl$_2$ through single-crystal XRD experiments.
The details of the experiment and the refined structure are described in the Methods section.
The difference from the previously reported structure is the lack of the $a$-glide, which leads to the space group $Cmc2_1$.
Figure~\ref{model}a illustrates intradimer and interdimer interactions expected from the crystal symmetry.
Although twofold rotation is absent, because of which two nearest-neighbor Cu atoms become symmetrically inequivalent, all the intradimer interactions remain identical.
On the other hand, the lack of the glide symmetry enables an alternation of interdimer interactions along the $a$-axis, while those along the $b$-axis remain uniform.
Finally, as shown in Fig.~\ref{model}b, three different hopping amplitudes can be present:
$J^A$, $J^{A\prime}$, and $J^B$, representing $\frac{1}{4}(J^a_{11} + J^a_{22} - J^a_{12} - J^a_{21}) $, $\frac{1}{4}(J^{a\prime}_{11} + J^{a\prime}_{22} - J^{a\prime}_{12} - J^{a\prime}_{21}) $,
and $\frac{1}{4}(J^b_{11} + J^b_{22} - J^b_{12} - J^b_{21}) $, respectively.

First, we discuss inelastic neutron scattering intensities sliced along the $H$ (or $K$) direction, which are shown as color contour maps in Fig.~\ref{dispersion1}a--c.
Intensity is integrated over the observed $L$ range to obtain good statistics.
At least two dispersive branches are clearly observed at 2--3~meV because of mixed domains:
they correspond to the same triplon band which is dispersive along both $H$ and $K$ directions.
In addition, the band exhibits the minimum energy at (2$m$, 2$n$, 0) ($m$, $n$: integer), indicating that triplon propagation is in-phase.
The three hopping amplitudes are all negative because of dominant antiferromagnetic diagonal interactions $J^\alpha_{12}$, $J^\alpha_{21}$,  $J^{a\prime}_{12}$, and $J^{a\prime}_{21}$, which is consistent with the results of DFT calculations~\cite{Okada}.

\begin{figure}[h]
\includegraphics[width=8cm, clip]{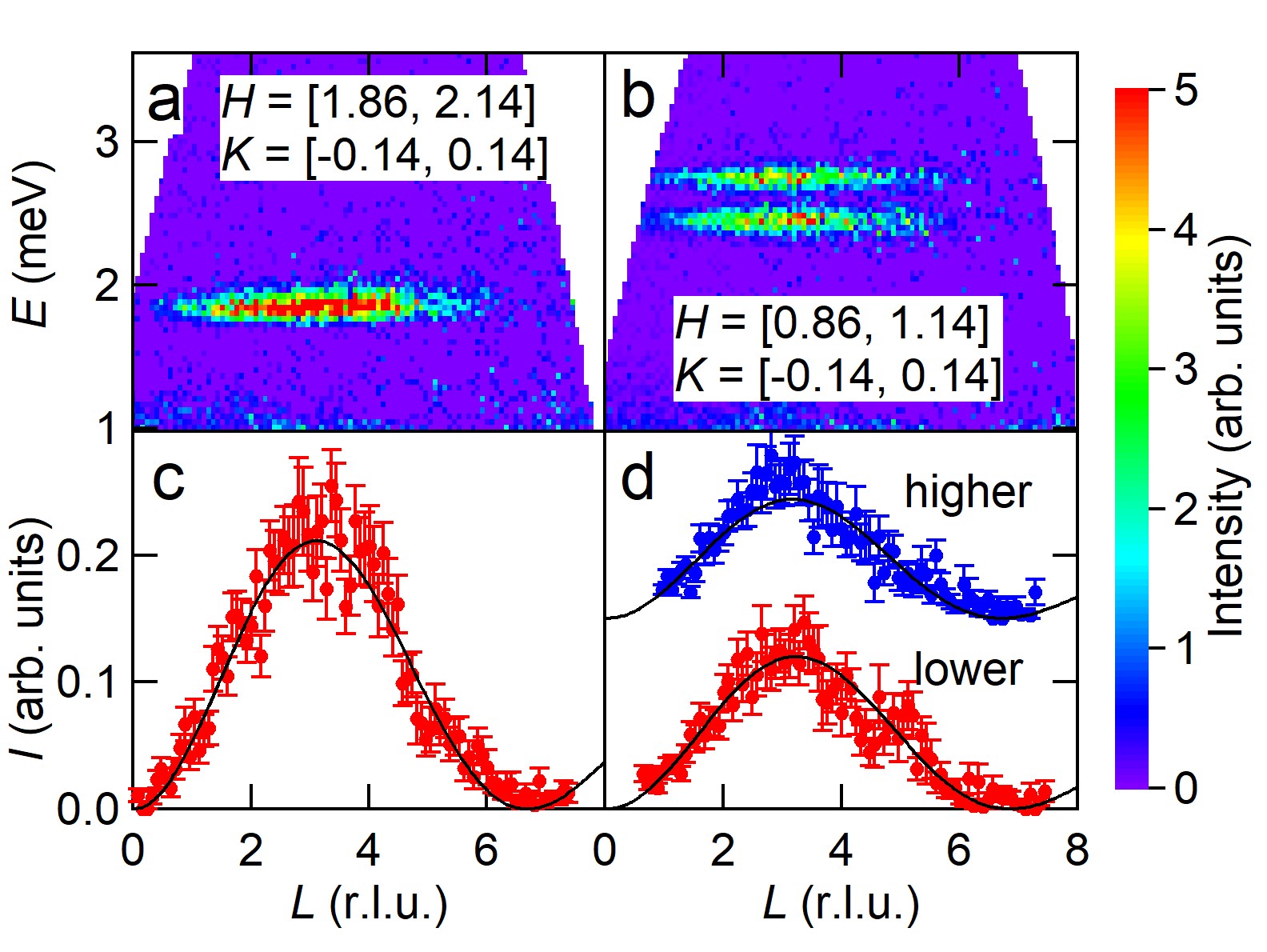} %300dpi
\caption{\label{dispersion2} {\bf Excitation spectra along L directions.}
{\bf a, b.} Color contour maps of scattering intensities along the $L$ direction.
Integrated $H$ and $K$ ranges are shown in the figures.
{\bf c, d} Integrated intensity plotted against $L$ together with a fit using the dimer structure factor represented by a solid black curve.
Red circles in Fig.~\ref{dispersion2}b are obtained by integrating the intensities shown in Fig.~\ref{dispersion2}a between 1.66 and 2.06~meV.
Red circles and blue squares in Fig.~\ref{dispersion2}d correspond to integrated intensities between 2.22 and 2.58 meV and between 2.62 and 2.98~meV from Fig.~\ref{dispersion2}c, respectively.
The intensity of the higher branch is shifted upward for clarity.}
\end{figure}

The contour maps of inelastic neutron scattering sliced along the $L$ direction are shown in Fig.~\ref{dispersion2}.
Figures~\ref{dispersion2}a and b represent integrated intensities around ($H$, $K$) =~(2, 0) (and (0, 2) from different domains) and (1, -1) (and (-1, 1)), respectively.
The excitations along $L$ is dispersionless, irrespective of $H$ and $K$, indicating good two-dimensionality in the exchange network.
In addition, integrated intensities are modulated along $L$, as should be the case with antiferromagnetically coupled dimers along the $c$-axis~\cite{Sasago}.
Figures~\ref{dispersion2}c and d show integrated intensities from Fig.~\ref{dispersion2}a and c, respectively.
The intensities of dimer antiferromagnets are often characterized by a dimer structure factor $I(Q, \omega) \sim~|f(Q)|^2 (1 - \cos (\mathbf{Q} \cdot \boldsymbol{r}_\mathrm{d})$),
where $f(Q)$ and $\boldsymbol{r}_\mathrm{d}$ indicate a form factor of Cu$^\mathrm{2+}$ and a vector representing intradimer separation, respectively.
Precisely speaking, the structure factor should be corrected if a few dimers with different orientations are present in a unit cell.
In Ba$_2$CuSi$_2$O$_6$Cl$_2$, there are four types of dimers with slightly different orientations.
However, since their canting angle of 0.9$^\circ$ from the $c$-axis is very small, we approximate that all the four dimers are aligned along the $c$-axis.
As shown in Fig.~\ref{dispersion2}c, the fit to this equation yields an $\boldsymbol{r}_\mathrm{d}$ of 0.150(1)$\mathbf{c}$,
which is consistent with 0.148(1)$\mathbf{c}$ obtained from the crystal structure.
The modulation along $L$ does not depend on $H$ and $K$, indicating that the whole intensities are from the equivalent dimer (Fig.~\ref{dispersion2}d).

%\begin{figure}[t]
%\includegraphics[width=8cm, clip]{energy_slice.jpg} %300dpi
%\caption{\label{eslice} {\bf Color contour maps of scattering intensities sliced with a constant energy of (a) 2.48, (b) 2.60, and (c) 2.72~meV.}
%Intensity is integrated for $\Delta E$ =~$\pm$0.06~meV and all measured $L$ values.
%White solid curves indicate boundaries between zero and nonzero intensity regions expected from the dispersion relation.}
%\end{figure}

What is \textit{not} expected for a simple dimer antiferromagnet is the decrease in intensity observed at 2.6~meV (Fig.~\ref{dispersion1} and Fig.~\ref{dispersion2}b),
which is almost independent of the scattering wave vector. %as shown in Fig.~\ref{eslice} by contour maps of intensities sliced with a constant energy.
%Apparently, the energy slice at 2.60~meV (Fig.~\ref{eslice}b) exhibits much weaker intensities than those at 2.48 (Fig.~\ref{eslice}a) and 2.72~meV (Fig.~\ref{eslice}c).
Note that the intensity decrease at a certain energy is not due to an extrinsic effect, because it is unchanged under different conditions (different $E_i$ and temperature, see Extended Fig.~\ref{fig:S3}).
The detailed $Q$ dependence of the triplon bands are shown in Fig.~\ref{dispersion1}f, which represents $Q$ slices of Fig.~\ref{dispersion1}a as functions of $E$.
At small $H$ and $K$, two peaks in $H$ and $K$ directions are overlapped with each other.
At 0.68 r.l.u., they separate into two. The right peak at 2.52~meV decreases and disappears above 0.80 r.l.u., while the left peak becomes more prominent with increasing $H$ and $K$.
Above 0.68~r.l.u., the new peak grows at 2.68~meV.

The coexistence of three modes at the same slice strongly indicates the presence of two triplon bands, which is not allowed when interdimer interactions along both $a$- and $b$-axes are uniform.
Let us start from the uniform case and then introduce the alternation to explain this phenomenon.
Note that triplon bands are degenerate since the crystallographic unit cell includes eight dimers, which are connected by a mirror symmetry with respect to the $bc$-plane, centering symmetry, and twofold screw symmetry along the $c$-axis.
For a simplicity, we do not count the fourfold degeneracy caused by the latter two symmetries and instead focus on two dimers in a primitive unit cell with a single layer, as shown in Fig.~\ref{model}b.
In the former case, only one continuous triplon band is detectable, and the structure factor of the other is almost 0.
The dispersion branches with strong and weak scattering intensities along the ($H$, 0) and (0, $K$) directions are depicted as solid and dashed curves in Fig.~\ref{dispersion1}d, respectively.
The high-energy band is not observable, because of a very small structure factor caused by two dimers aligned to almost the same direction.
In other words, the dimer orientations are so close to each other that triplet excitations cannot be distinguished from those expected from a hypothetical unit cell including one dimer. 
The presence of a triplon band with very weak intensities was also reported in TlCuCl$_3$~\cite{Oosawa}.

When the alternation along the $a$-direction is introduced,
a band inversion between the low- and high- energy bands induces a gap between them, as shown in Fig.~\ref{dispersion1}e.
If the alternation is very small, the structure factor becomes very close to that represented in Fig.~\ref{dispersion1}d.
Thus, the intensities of the low- and high- energy bands greatly vary around a crossing point of 0.74~r.l.u., indicated by arrows in Figs.~\ref{dispersion1}d and e;
the intensities of the high-energy band significantly increases above the crossing point, while those of the low-energy band become undetectable.
Even at different $H$ and $K$, the band crossing occurs at the same energy, $J$, since the two dimers are symmetrically equivalent.
Consequently, the wavevector-independent gap centered at $J$ appears between two triplon bands, as we have discussed.
Note that the alternation is only allowed along the $a$-axis owing to the symmetry.
This enables the indexing of all the excitations, as denoted in Fig.~\ref{dispersion1}a--c.

Next, we derive an analytical form of the dispersion relation to determine the exchange constants and then discuss a topological character of the gap between the two triplon bands.
For this purpose, a bond-operator approach~\cite{bd} is applied to the 2D dimer model represented in Fig.~\ref{model}a.
Creation operators $s^\dagger_{mn}$ representing a singlet state and $t^\dagger_{xmn}$, $t^\dagger_{ymn}$, and $t^\dagger_{zmn}$ representing triplet states are introduced,
%Triplon bond operators representing a singlet state and triplet states are defined as
%$s^\dagger_{mn} |0 \rangle =~\frac{1}{\sqrt{2}} (|\uparrow \rangle_{mn1} |\downarrow \rangle_{mn2} - |\downarrow \rangle_{mn1} |\uparrow \rangle_{mn2})$, 
%$t^\dagger_{xmn} |0 \rangle =~-\frac{1}{\sqrt{2}}~(|\uparrow \rangle_{mn1} |\uparrow \rangle_{mn2} - |\downarrow \rangle_{mn1} |\downarrow \rangle_{mn2})$, 
%$t^\dagger_{ymn} |0 \rangle =~\frac{i}{\sqrt{2}}~(|\uparrow \rangle_{mn1} |\uparrow \rangle_{mn2} + |\downarrow \rangle_{mn1} |\downarrow \rangle_{mn2})$, and
%$t^\dagger_{zmn} |0 \rangle =~\frac{1}{\sqrt{2}} (|\uparrow \rangle_{mn1} |\downarrow \rangle_{mn2} + |\downarrow \rangle_{mn1} |\uparrow \rangle_{mn2})$, respectively,
where $m$ and $n$ are labels used to distinguish dimers, and $1$ and $2$ indicate two Cu atoms in a single dimer.
%The above definition leads to interacting hard-core bosons characterized by hopping amplitudes $J^A$, $J^{A\prime}$, and $J^B$, as depicted in Fig.~\ref{model}b.
Then the Hamiltonian can be reduced to interacting hard-core bosons characterized by hopping amplitudes $J^A$, $J^{A\prime}$, and $J^B$, as depicted in Fig.~\ref{model}b.
The detailed calculations are described in the Methods section.
A $k$-dependent form of a Hamiltonian is obtained by Fourier transformation as
\begin{equation}
{\cal H} \sim \frac{1}{2} \sum_\mathbf{k} \sum_\alpha 
{\cal T}^\dagger {\cal M}_\mathbf{k} {\cal T},
\label{matrix}
\end{equation}
where
\begin{equation}
{\cal T} = \left( \begin{array}{c} t^1_{\alpha\mathbf{k}} \\ t^2_{\alpha\mathbf{k}} \\ t^{\dagger, 1}_{\alpha(-\mathbf{k})} \\ t^{\dagger, 2}_{\alpha(-\mathbf{k})} \end{array} \right), 
{\cal M}_\mathbf{k} =~\left( ~\begin{matrix} J & \Lambda_\mathbf{k} & 0 & \Lambda_\mathbf{k} \\ \Lambda^*_\mathbf{k} & J & \Lambda^*_\mathbf{k} & 0 \\
0 & \Lambda_\mathbf{k} & J & \Lambda_\mathbf{k} \\ \Lambda^*_\mathbf{k} & 0 & \Lambda^*_\mathbf{k} & J \end{matrix}  \right),
\label{matrix2}
\end{equation}
and $\Lambda_\mathbf{k} = J^A e^{-i k_x a/2} + J^{A\prime} e^{i k_x a/2} + J^B \big( e^{-i k_y b/2} + e^{i k_y b/2} \big)$.
The superscripts on each operator denote the two sublattices in the primitive unit cell.
Quadratic terms from $\alpha =~x, y, z$ are block-diagonalized into the same matrix, ${\cal M}_\mathbf{k}$, reflecting that each band is triply degenerate owing to a rotation symmetry.
Dispersion relations are obtained by applying Bogoliubov transformation: by diagonalizing the matrix $\boldsymbol{\Sigma} {\cal M}_\mathbf{k}$ ($\boldsymbol{\Sigma} = \mathrm{diag}(1,1,-1,-1)$),
dispersion relations are obtained as
\begin{equation}
E_{\pm,\mathbf{k}} = \sqrt{J^2 \pm 2 J |\Lambda_\mathbf{k}|^2}.
\label{diseq}
\end{equation}

The observed triplon bands are well reproduced by the dispersion relation given by eq.~\eqref{diseq}.
The two bands with a large and small structure factor are represented by thick solid and thin dashed curves in Fig.~\ref{dispersion1}a--c, respectively.
The parameters $J$, $J^A$, $J^{A\prime}$, and $J^B$ are selected as 2.61, $-$0.24, $-$0.16, and $-$0.13~meV, respectively, because these values best reproduce the observed dispersions.
The simulated dispersion curves perfectly agree with the observed bands.
This model is also supported by the energy slice presented in Extended~Fig.~\ref{eslice}.
%The dashed curves in the figure indicate the region where triplon bands cross with a constant energy; an energy width of $\Delta E =~\pm0.10$ is taken into account from the energy window and energy resolution.
%They well describe the area where finite intensities are observed.
%Even around the gap energy, 2.60$\pm$0.06~meV, weak intensities are detected because of the narrow band gap.
These parameters are also consistent with $J$ =~2.4~meV and $|J_p - J_d|$ (equal to $|J^A + J^{A\prime} + 2 J^B|/2)$ =~0.30~meV estimated from the magnetization curve~\cite{Okada}.

\begin{figure}[t]
\includegraphics[width=8cm, clip]{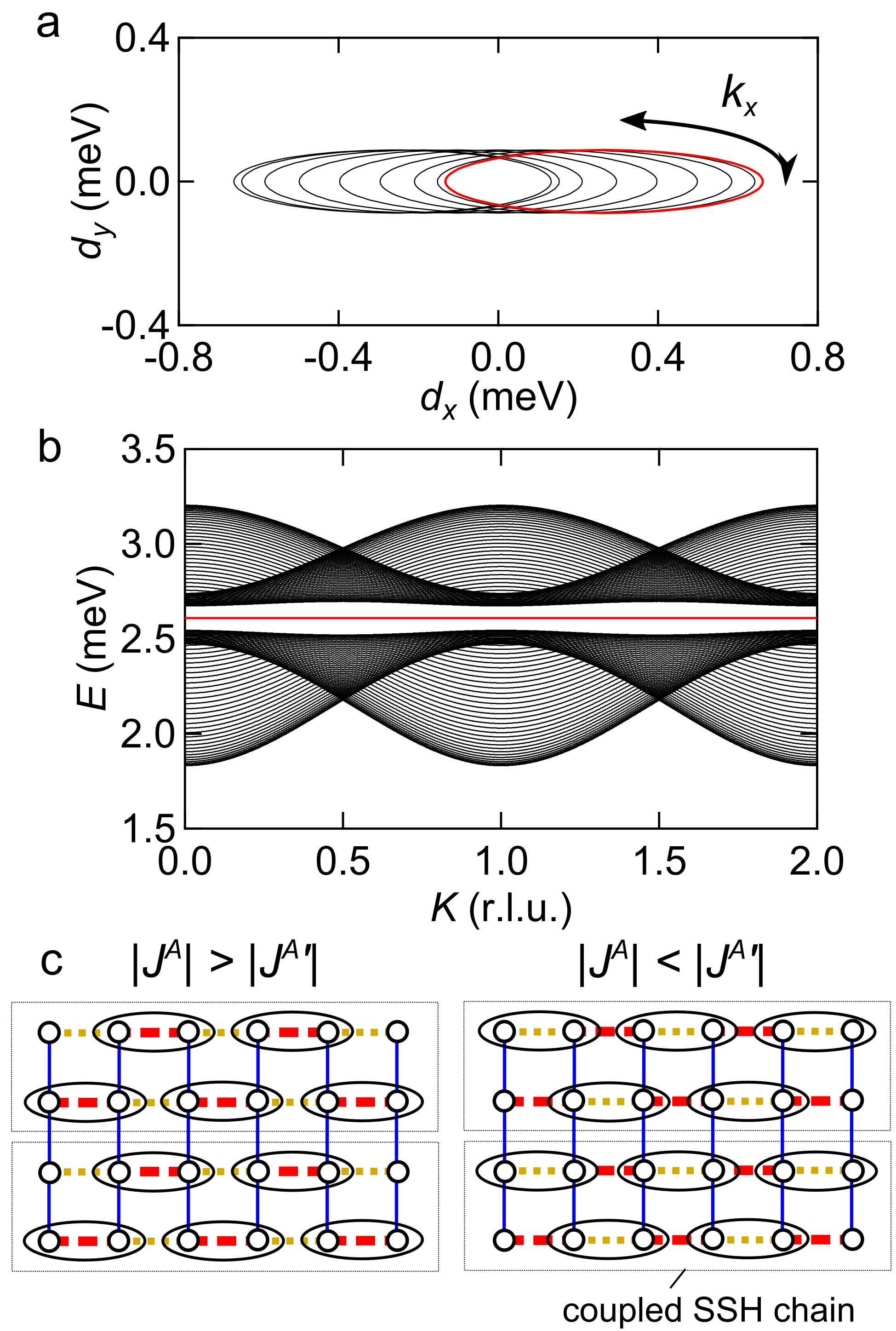} %less than 300 dpi
\caption{\label{edge} {\bf Schematic view of expected pseudomagnetic field and edge states.} 
{\bf a.} Distribution of a pseudomagnetic field induced on the triplon bands. Each solid curve represents a pseudomagnetic field with a fixed $k_y$.
{\bf b.} Energy spectrum along the $K$ direction. The calculation is based on the model presented in Fig.~\ref{model}b, in which open boundary conditions with 100 sites (for each sublattice) are imposed along the $a$ direction while periodic boundary conditions are set along the $b$ direction.
Half of the all modes are shown, and edge states are stressed by a red line for clarity.
{\bf c.} The interdimer network in Ba$_2$CuSi$_2$O$_6$Cl$_2$ shown as SSH chains coupled by interchain hoppings. The same symbols and lines as those in Fig.~\ref{model} are used.
An ellipsoid and a dashed rectangle represent a pair of dimers with a larger hopping amplitudes and a single coupled SSH chain, respectively.}
\end{figure}

Interestingly, the gap between two triplet bands is topologically nontrivial.
This can be easily understood by neglecting pair creation and annihilation terms in eq.~\eqref{matrix}, which do not alter topological properties as we discuss later.
Equation~\eqref{matrix} is now reduced to a simple form:
\begin{equation}
{\cal H} \sim \sum_\mathbf{k} \left( t^{\dagger, 1}_{+, \mathbf{k}} t^{\dagger, 2}_{+, \mathbf{k}} \right)
{\cal M}^\prime_\mathbf{k}
\left( \begin{array}{c} t^1_{+, \mathbf{k}} \\ t^2_{+, \mathbf{k}} \end{array} \right),
\end{equation}
with a 2 $\times$ 2 matrix
\begin{equation}
{\cal M}^\prime_\mathbf{k} = \left( \begin{matrix} J & \Lambda_\mathbf{k} \\ \Lambda^*_\mathbf{k} & J \end{matrix} \right) = J \boldsymbol{1} + \boldsymbol{d} \cdot \boldsymbol{\sigma},
\label{easymatrix}
\end{equation}
where $\boldsymbol{d}$ and $\boldsymbol{\sigma}$ represent a pseudomagnetic field, $\boldsymbol{d} = \left( \mathrm{Re} \Lambda_\mathbf{k}, -\mathrm{Im} \Lambda_\mathbf{k}, 0 \right)$
and a Pauli matrix, respectively.
The matrix leads to the eigenenergy $E_\mathbf{k} = J \pm |\boldsymbol{d}| = J \pm |\Lambda_\mathbf{k}|$.
Thus, an energy gap exists between the two modes if $|\Lambda_\mathbf{k}| > 0$ holds for all $\mathbf{k}$; $J^A \neq J^{A\prime}$ and $|J^A +J^{A\prime}|/2 >~ |J^B|$.

The matrix ${\cal M}^\prime_\mathbf{k}$ represents a quasi-one-dimensional (quasi-1D) extension of the Su-Schrieffer-Heeger (SSH) model~\cite{SSH, SSH2}.
The SSH model describes electron motions in a 1D lattice with alternating hopping amplitudes and well demonstrates a topological distinction between nontrivial and trivial phases with respect to the number of edge states.
Even for bosonic systems such as triplons, the same topological distinction can be made between excited modes if an energy gap exists between them.
The hopping amplitudes of triplons in Ba$_2$CuSi$_2$O$_6$Cl$_2$ are alternated along the $a$-direction but uniform along the $b$-direction, as shown in Fig.~\ref{model}b.
Thus, the interdimer network in Ba$_2$CuSi$_2$O$_6$Cl$_2$ can be regarded as SSH chains \textit{coupled by interchain hoppings}.
Under $|J^A +J^{A\prime}|/2 >~ |J^B|$, the variation of $k_y$ only causes a small shift of $\boldsymbol{d}$ along $d_x$, keeping the winding number unchanged.
Thus, the winding number can be defined for a fixed $k_y$, as that defined for the 1D-system.
%The matrix ${\cal M}^\prime_\mathbf{k}$ represents a quasi-one-dimensional (quasi-1D) extension of the Su-Schrieffer-Heeger (SSH) model~\cite{SSH, SSH2}.
%The SSH model describes electron motions in a 1D lattice with alternating hopping amplitudes and well demonstrates a topological distinction between nontrivial and trivial phases with respect to the number of edge states.
%Even for bosonic systems such as triplons, the same topological distinction can be made between excited modes if an energy gap exists between them.
%The hopping amplitudes of triplons in Ba$_2$CuSi$_2$O$_6$Cl$_2$ are alternated along the $a$-direction but uniform along the $b$-direction, as shown in Fig.~\ref{model}b.
%Thus, the interdimer network can be regarded as SSH chains coupled by uniform interchain interactions.

It should be noted that the alternating sequence of intrachain hopping amplitudes is opposite between nearest-neighbor SSH chains.
The alternation yields two nontrivial gapped phases with changing intrachain hopping amplitudes~\cite{coupledSSH}, while the SSH chain yields one trivial and one nontrivial phases~\cite{SSH, SSH2}.
The winding number is evaluated by projecting a pseudomagnetic field $\boldsymbol{d}$ on $d_x$-$d_y$ space.
Figure~\ref{edge}a depicts $\boldsymbol{d}$ with exchange parameters set to those determined from the present experiment.
For a fixed $k_y$, $\boldsymbol{d}$ represents a single ellipsis, as well as that of the SSH chain.
The winding number can be defined as the number with which $\boldsymbol{d}$ surrounds the origin counterclockwise or clockwise; $N = \pm1$ for the present case.
The two phases with the opposite winding number are separated by a phase boundary at $J^A = J^{A\prime}$, where the gap is closed.
The winding number cannot be changed without closing the gap, because of a chiral symmetry ($d_z = 0$).

The above discussion indicates that edge states protected by a chiral symmetry exist in the triplon band gap observed in Ba$_2$CuSi$_2$O$_6$Cl$_2$ as well as the SSH model.
The symmetry-protected edge states cannot be removed by pair creation and annihilation terms, which can be confirmed by deriving the Berry connection. 
A Bogoliubov-de Gennes form of the two-sublattice triplon-band Hamiltonian is generalized into a 4$\times$4 matrix as follows:
\begin{equation}
{\cal M}^\mathrm{(4)}_\mathbf{k} = \left( \begin{matrix} J\mathbf{1}+\boldsymbol{d}\cdot\boldsymbol{\sigma} & \boldsymbol{d}\cdot\boldsymbol{\sigma} \\ \boldsymbol{d}\cdot\boldsymbol{\sigma} & J\mathbf{1}+\boldsymbol{d}\cdot\boldsymbol{\sigma} \end{matrix} \right),
\label{gH}
\end{equation}
where $\boldsymbol{d} = (d_{x}, d_{y}, d_{z})$ is a three-component real vector that is a function of $\mathbf{k}$.
%Then, by following the definition of the Berry connection for bosonic systems \cite{magnon1, MagnonHall3}, its real part is obtained as
%\begin{equation}
%\mathrm{Re} A_{\pm,\mathbf{k}} =~\pm \frac{1}{2|\boldsymbol{d}| (|\boldsymbol{d}| + d_z)} \left( d_x \frac{\partial d_y}{\partial k_x} - d_y \frac{\partial d_x}{\partial k_x} \right),
%\label{Re}
%\end{equation}
%where $\pm$ represents two subbands.
Irrespective of which gauge is selected, the real part of the Berry connection corresponds to that derived from a 2$\times$2 matrix, ${\cal M}^\mathrm{(2)}_\mathbf{k} = J\mathbf{1}+\boldsymbol{d}\cdot\boldsymbol{\sigma}$,
implying that topological properties are the same for both Hamiltonians, ${\cal M}^\mathrm{(4)}_\mathbf{k}$ and ${\cal M}^\mathrm{(2)}_\mathbf{k}$ (see the Methods section for a detailed derivation).
%In fact, for a 1D system in which $\boldsymbol{d}$ is dependent on $k_x$, $d_z = 0$ corresponds to the Zak phase \cite{Zak} $\gamma_{\pm} = - \int_{\mathrm{BZ}} dk_x A_{\pm,\mathbf{k}}$ quantized into $\pm n \pi$,
%where $n$ corresponds to the winding number, $\int_\mathrm{BZ} dk_x (d_x {\partial d_y}/{\partial k_x} - d_y {\partial d_x}/{\partial k_x})/(2 \pi |\boldsymbol{d}|^2)$.
This indicates that edge states are protected by the equivalence between the two sublattices, which corresponds to the chiral symmetry if pair creation and annihilation terms are absent.
%However, these terms make the equivalence different from the chiral symmetry because of the asymmetry between the two bands above and below $J$ (eq.~\eqref{diseq}).

The presence of edge states is also confirmed by calculating the energy spectrum with a finite length of chains.
As shown in Fig. \ref{edge}b, the twofold-degenerate edge states appear at the energy $J$ in addition to the bulk bands with dispersion relations described by eq.~\eqref{diseq}.
The flat dispersion reflects triplon densities localized at the edge: the alternation of hopping amplitudes induces one ''unpaired'' triplon at each edge, as illustrated in Fig. \ref{edge}c.
The winding number can be reversed by placing the ''unpaired'' triplon on the other sublattice, which can be realized by reversing the magnitudes of $J^A$ and $J^{A\prime}$.

It should be noted that the edge states in the present model are induced by a bipartite nature,
and edge states from the $S_z = 1, 0$ and $-$1 branches of triplet excitations are degenerate in the present model.
This is in contrast with the model based on SrCu$_2$(BO$_3$)$_2$ with both interdimer Dzyaloshinskii-Moriya interactions and a magnetic field~\cite{edgestate1, edgestate3}.
The experimental detection of edge states in the triplon band gap is a future task.

In summary, triplet excitations in the dimerized quantum magnet Ba$_2$CuSi$_2$O$_6$Cl$_2$ were investigated via inelastic neutron scattering experiments.
Two modes of triplet excitations were detected together with a clear energy gap, which is induced by alternation of the interdimer interactions along the $a$-axis.
%Two modes of triplet excitations were detected together with a clear energy gap
%This result is consistent with the newly determined crystal structure: the lack of $a$-glide allows interdimer interactions along the $a$-axis to alternate, while those along the $b$-axis become uniform.
The whole dispersion relations are well reproduced with the three hopping constants $J^A$, $J^{A\prime}$, and $J^B$.
The correspondence between the interdimer network of Ba$_2$CuSi$_2$O$_6$Cl$_2$ and a quasi-1D extension of the coupled SSH model indicates
topological protected edge states in the triplon band gap.

\begin{acknowledgments}
The neutron scattering experiment was performed under the J-PARC user program (Proposal No. 2016B0023).
We express our sincere gratitude to M. Matsumoto and K. Nomura for useful discussions and comments.
This work was supported by Grants-in-Aid for Scientific Research (A) (No.~17H01142), (C) (No.~16K05414), and Challenging Research （Exploratory） (No.~17K18744) from the Japan Society for the Promotion of Science.
\end{acknowledgments}

\section{Author contributions}
H. T. designed the experiment. K. T. and H. T. grew the crystal.
K. T., K. Nawa, N. K., H. T., S. O. -K. and K. Nakajima performed the INS experiments.  
K. Nawa and T. J. S. worked out the neutron-data and theoretical analysis.
H. S., K. Nawa, T. J. S., and H. U. performed the single crystal XRD experiments.
K. Nawa and H. T. wrote the manuscript. 

%\section{Data availability}
%The data that support the plots within this paper and other findings of this study are available from the corresponding authors upon reasonable request.
%
%\section{Competing interests}
%The authors declare no competing interests.

\bibliographystyle{naturemag}
\bibliography{article}% Produces the bibliography via BibTeX.
%\end{document}
\clearpage

\setcounter{figure}{0}
\makeatletter 
\renewcommand{\fnum@figure}{\textbf{Extended Figure~\thefigure}}
\makeatother
\makeatletter
\renewcommand{\thetable}{\arabic{table}}
\renewcommand{\fnum@table}{\textbf{Extended Table~\thetable}}
\makeatother

% ****** Start of file apssamp.tex ******
%
%   This file is part of the APS files in the REVTeX 4.1 distribution.
%   Version 4.1r of REVTeX, August 2010
%
%   Copyright (c) 2009, 2010 The American Physical Society.
%
%   See the REVTeX 4 README file for restrictions and more information.
%
% TeX'ing this file requires that you have AMS-LaTeX 2.0 installed
% as well as the rest of the prerequisites for REVTeX 4.1
%
% See the REVTeX 4 README file
% It also requires running BibTeX. The commands are as follows:
%
%  1)  latex apssamp.tex
%  2)  bibtex apssamp
%  3)  latex apssamp.tex
%  4)  latex apssamp.tex
%
%\documentclass[%
% reprint,
%superscriptaddress,
%groupedaddress,
%unsortedaddress,
%runinaddress,
%frontmatterverbose,
%preprint,
%showpacs,preprintnumbers,
%nofootinbib,
%nobibnotes,
%bibnotes,
% amsmath,amssymb,
% aps,
%prl
%pra,
%prb,
%rmp,
%prstab,
%prstper,
%floatfix,
%]{revtex4-1}

%\usepackage{graphicx}% Include figure files
%\usepackage{dcolumn}% Align table columns on decimal point
%\usepackage{bm}% bold math
%\usepackage{amsmath}
%\usepackage{hyperref}% add hypertext capabilities
%\usepackage[mathlines]{lineno}% Enable numbering of text and display math
%\usepackage{color}
%\linenumbers\relax % Commence numbering lines

%\usepackage[showframe,%Uncomment any one of the following lines to test
%%scale=0.7, marginratio={1:1, 2:3}, ignoreall,% default settings
%%text={7in,10in},centering,
%%margin=1.5in,
%%total={6.5in,8.75in}, top=1.2in, left=0.9in, includefoot,
%%height=10in,a5paper,hmargin={3cm,0.8in},
%]{geometry}

%\begin{document}

\section{\label{methods}Methods}
{\bf Sample preparation}
Single crystals of Ba$_2$CuSi$_2$O$_6$Cl$_2$ were synthesized according to the procedure described in Ref.~\cite{Okada}.
To synthesize single crystals of Ba$_2$CuSi$_2$O$_6$Cl$_2$, we first prepared Ba$_2$CuTeO$_6$ powder through a solid-state reaction. A mixture of Ba$_2$CuTeO$_6$ and BaCl$_2$ in a molar ratio of $1\,{:}\,10$ was vacuum-sealed in a quartz tube, which acts as a SiO$_2$ source. The temperature at the center of the horizontal tube furnace was lowered from 1100$^{\circ}$C to 800$^{\circ}$C over 10 days. Plate-shaped blue single crystals with a maximum size of $10\,{\times}\,10{\times}\,1.5$ mm$^3$ were obtained. The wide plane of the crystals was confirmed to be the crystallographic $ab$ plane by X-ray diffraction. The quartz tube frequently exploded during cooling to room temperature after the crystallization process from 1100$^{\circ}$C to 850$^{\circ}$C. To avoid hazardous conditions and damage to the furnace, a cylindrical nichrome protector was inserted in the furnace core tube.

{\bf Single-crystal x-ray diffraction experiments}
Because the band gap of triplet excitations observed in \\
Ba$_2$CuSi$_2$O$_6$Cl$_2$ cannot be described by the exchange model based on the original crystal structure~\cite{Okada}, we reexamined the crystal structure at room temperature by using a RIGAKU R-AXIS RAPID three-circle X-ray diffractometer equipped with an imaging plate area detector. Monochromatic Mo-K$\alpha$ radiation with a wavelength of ${\lambda}\,{=}\,0.71075$\,\rm{\AA} was used as the X-ray source. Data integration and global-cell refinements were performed using data in the range of $3.119^{\circ}\,{<}\,{\theta}\,{<}\,30.508^{\circ}$, and absorption correction based on face indexing and integration on a Gaussian grid was also performed. The total number of reflections observed was 73781, among which 5947 reflections were found to be independent and 5096 reflections were determined to satisfy the criterion $I\,{>}\,2{\sigma}(I)$. Structural parameters were refined by the full-matrix least-squares method using SHELXL-97 software. The final $R$ indices obtained for $I\,{>}\,2{\sigma}(I)$ were $R\,{=}\,0.0376$ and $wR\,{=}\,0.0803$. The crystal data are listed in Extended Table \ref{table:1}. The structure of Ba$_2$CuSi$_2$O$_6$Cl$_2$ is orthorhombic $Cmc2_1$ with cell dimensions of $a\,{=}\,13.9064(3)$\,$\rm{\AA}$, $b\,{=}\,13.8566(3)$\,$\rm{\AA}$, $c\,{=}\,19.5767(4)$\,$\rm{\AA}$, and $Z\,{=}\,16$. Its atomic coordinates and equivalent isotropic displacement parameters  are shown in Extended Table \ref{table:2}.

\begin{figure}[t]
\begin{center}
\includegraphics[width=8.2cm,clip]{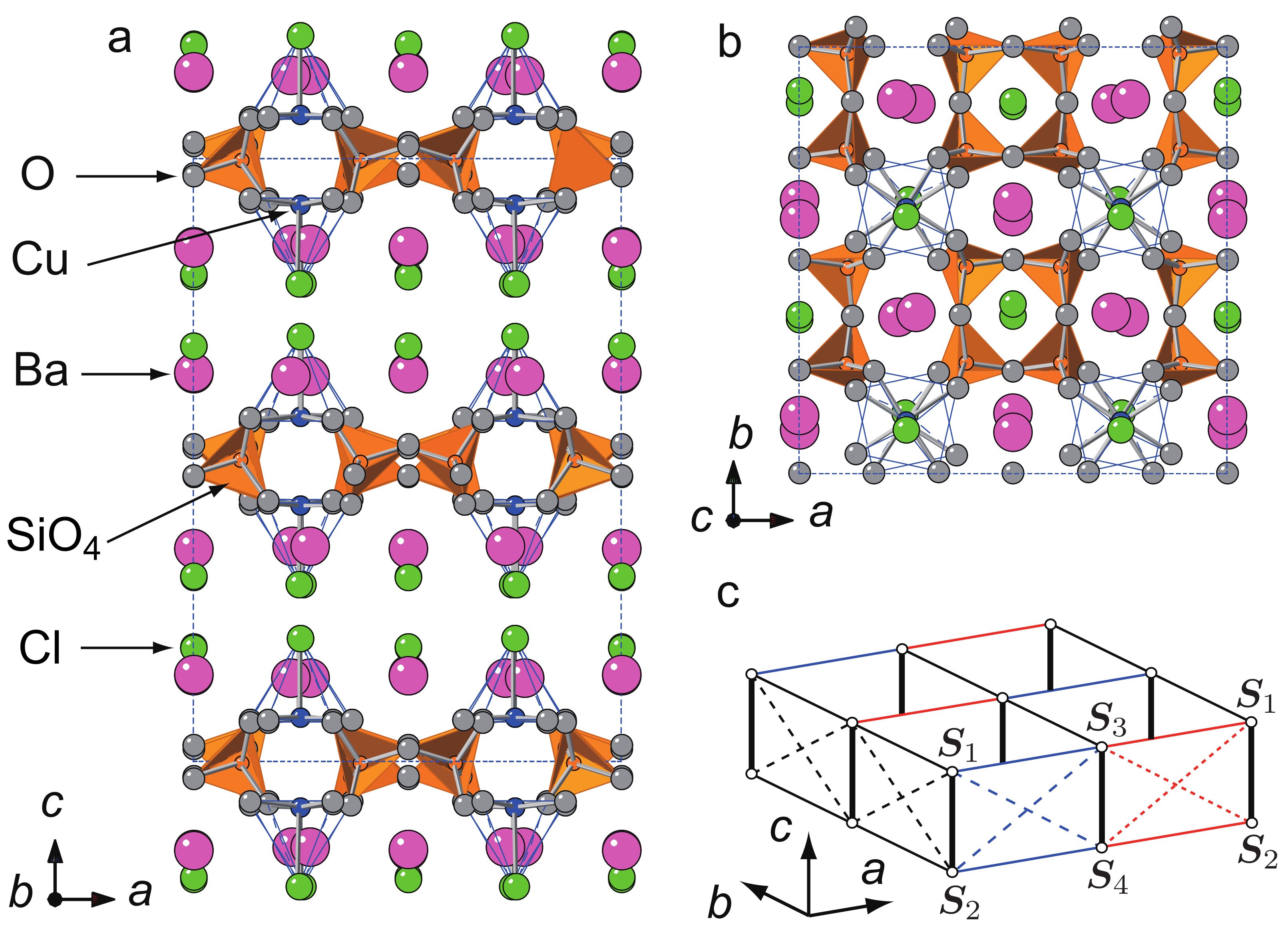}
\caption{{\bf a,b.} Crystal structure of Ba$_2$CuSi$_2$O$_6$Cl$_2$ viewed along ({\bf a}) the $b$-axis and ({\bf b}) the $c$-axes. Dashed lines indicate a unit cell. {\bf c.} 2D model of the exchange network. Thick solid lines represent the intradimer exchange interaction $J$. Thin solid, dashed, and dotted lines represent the interdimer exchange interactions, which are alternating along the $a$ axis. }
\label{fig:structure}
\end{center}
\end{figure}

Extended Figure~\ref{fig:structure} shows the redetermined crystal structure of Ba$_2$CuSi$_2$O$_6$Cl$_2$. The structure is closely related to that of Ba$_2$CoSi$_2$O$_6$Cl$_2$~\cite{Tanaka}. The crystal structure has a CuO$_4$Cl pyramid feature with a Cl$^-$ ion on an apex. The CuO$_4$Cl pyramids are linked via SiO$_4$ tetrahedra in the $ab$ plane. Magnetic spin-1/2 Cu$^{2+}$ is located at the center of the base composed of O$^{2-}$, which is parallel to the $ab$
plane. Two neighboring CuO$_4$Cl pyramids along the $c$ axis are
placed with their bases facing each other. The CuO$_4$Cl pyramids are
linked via SiO$_4$ tetrahedra in the $ab$ plane. The atomic linkage in the $ab$ plane
is approximately the same as that of
BaCuSi$_2$O$_6$~\cite{Sparta,Sasago}.

It is natural to assume from the crystal structure that two Cu$^{2+}$ spins located on the bases of neighboring CuO$_4$Cl pyramids along the $c$ axis form an antiferromagnetic dimer,
and the dimers are coupled by weak exchange interactions in the $ab$ plane.
In fact, the presented excitation spectrum supports this model. 
The exchange network of Ba$_2$CuSi$_2$O$_6$Cl$_2$ is illustrated in Extended Fig.~\ref{fig:structure}c. 
In the original crystal structure reported in Ref.~\cite{Okada}, there is no alternation of the interdimer interactions along the $a$ and $b$ axes, while in the redetermined structure the interdimer interactions are alternate along the $a$ axis.
It is also close to a 2D exchange network in BaCuSi$_2$O$_6$~\cite{Sebastian, Sasago}.
However, it should be emphasized that all the dimers are symmetrically equivalent in Ba$_2$CuSi$_2$O$_6$Cl$_2$,
while three inequivalent dimers are resolved in BaCuSi$_2$O$_6$ owing to a structural transition~\cite{Samulon2}.

\begin{figure}[h]
\includegraphics[width=5.0cm, clip]{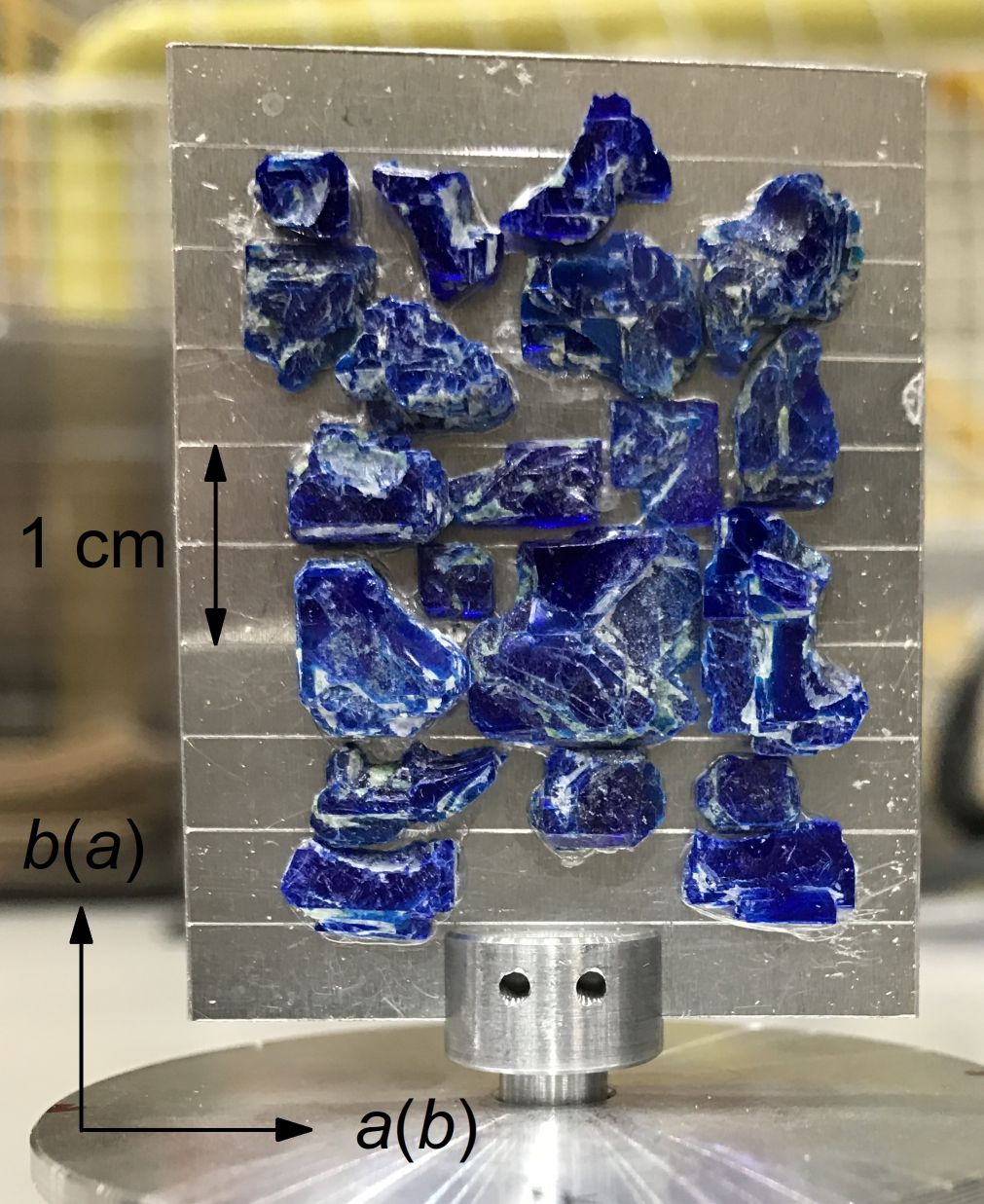}
\caption{Ba$_2$CuSi$_2$O$_6$Cl$_2$ crystals co-aligned on an aluminum plate.}
\label{fig:S1}
\end{figure}

{\bf Inelastic neutron scattering experiments}
To explore the 2D nature of triplon excitations in Ba$_2$CuSi$_2$O$_6$Cl$_2$, its magnetic excitations were investigated
using the cold-neutron disk chopper spectrometer AMATERAS installed in the Materials and Life Science Experimental Facility at J-PARC, Japan~\cite{AMATERAS}.
As shown in Extended Figure~\ref{fig:S1}, twenty pieces of single crystals were coaligned on a rectangular Al plate so that an $a^*$- or $b^*$- direction for every crystal coincided with the edge directions of the Al plate.
The Al plate was fixed in a vertical direction to set the $a^*$- and $c^*$-axes or  $b^*$- and $c^*$-axes in the horizontal plane.
Note that $a^*$- and $b^*$-axes cannot be distinguished with each other because of crystallographic domains.
Thus, both $a^*$- and $b^*$- components of a scattering vector are converted to a reciprocal lattice unit by the average of $a$ and $b$-axis lengths, which is 13.88~\AA.
The mixed domains do not matter in our analysis, because $a$- and $b$-axis lengths are almost the same,
and triplet bands along $a^*$ and $b^*$ directions can be easily distinguished with each other, as described in the main text.
Incident neutron energies were set to $E_i$ = (23.65, 5.924)  meV and (7.732, 3.135)~meV
by using repetition multiplication~\cite{Nakamura}.
The coaligned crystals were rotated between a direction that forms a bond angle of $-$35$^\circ$ and 55$^\circ$ with respect to the $c^*$-axis for $E_i$ = (23.65, 5.924) meV,
while incident neutrons were kept parallel to the $c^*$-axis for $E_i$ = (7.732, 3.135)~meV.
The sample was cooled down to 0.3 and 2.5~K by using a $^3$He refrigerator.
All the data collected were analyzed using the software suite UTSUSEMI~\cite{UTSUSEMI}.

\begin{figure}[h]
\includegraphics[width=8cm, clip]{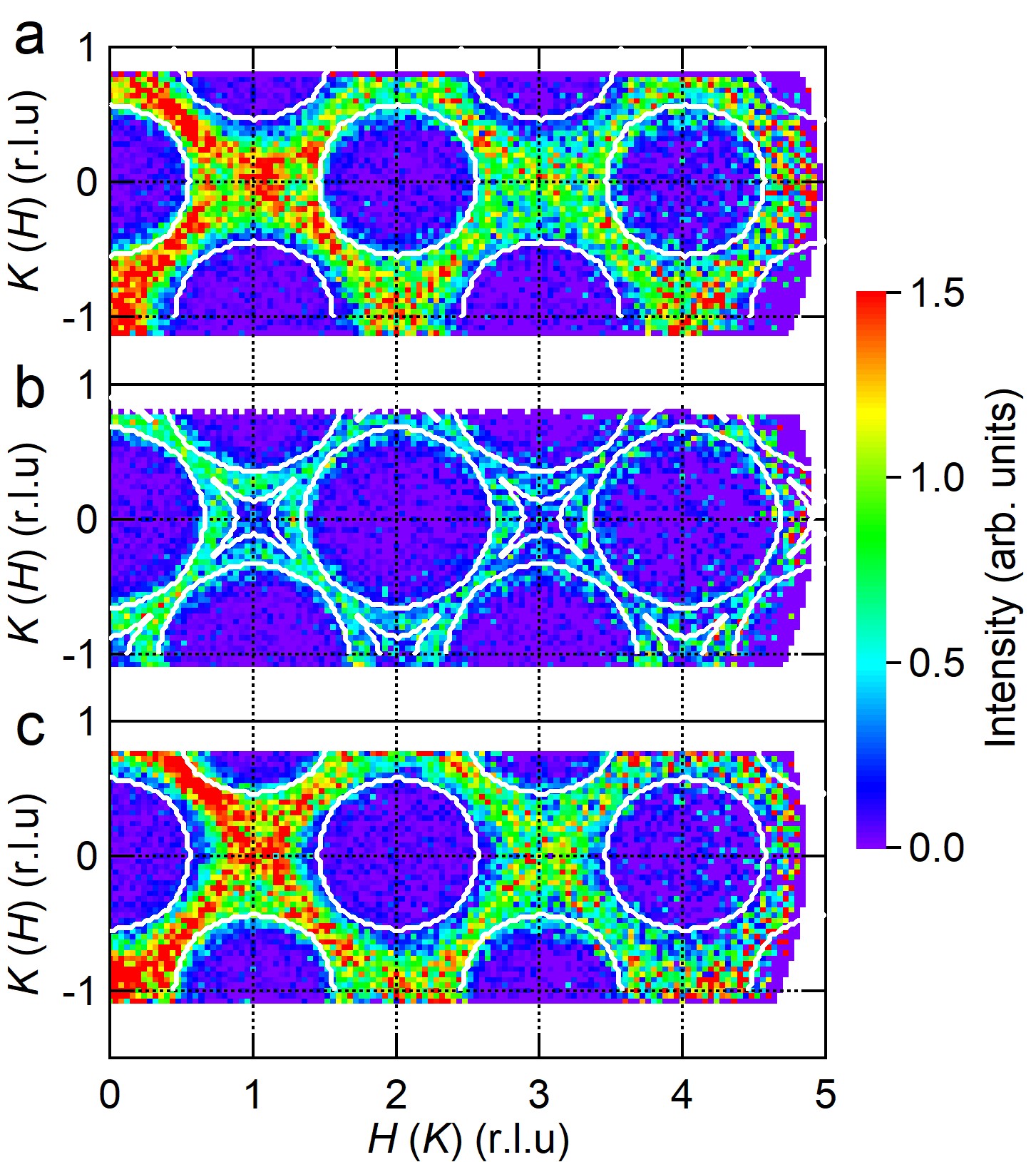} %300dpi
\caption{\label{eslice} {\bf Color contour maps of scattering intensities sliced with a constant energy of (a) 2.48, (b) 2.60, and (c) 2.72~meV.}
Intensity is integrated for $\Delta E$ =~$\pm$0.06~meV and all measured $L$ values.
White solid curves indicate boundaries between zero and nonzero intensity regions expected from the dispersion relation.}
\end{figure}

The model presented in the main text is also supported by the energy slice presented in Extended Fig.~\ref{eslice}.
The dashed curves in the figure indicate the region where triplon bands cross with a constant energy; an energy width of $\Delta E =~\pm0.10$ is taken into account from the energy window and energy resolution.
They well describe the area where finite intensities are observed.
Even around the gap energy, 2.60$\pm$0.06~meV, weak intensities are detected because of the narrow band gap.

\begin{figure}[h]
\includegraphics[width=7.0cm, clip]{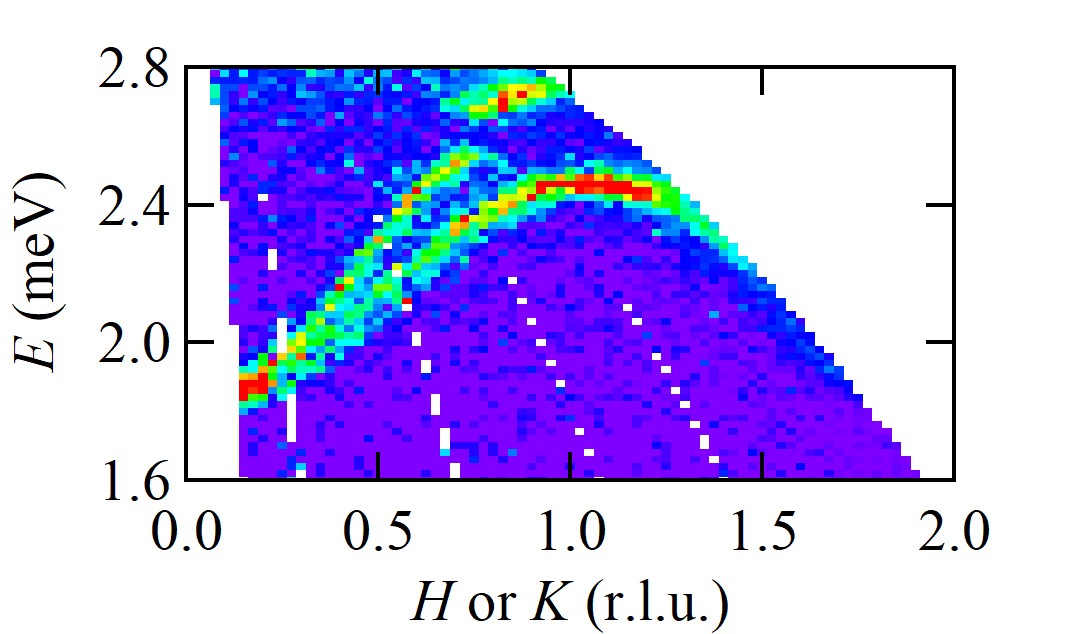}
\caption{Color contour maps of the scattering intensities sliced along the ($H$, 0) and (0, $K$) direction.
Intensity is integrated along both $K$ with a thickness of $\Delta K$ =~$\pm$0.06 and all collected $L$ values.
The presented data are collected at 2.5~K with an $E_i$ of 3.14 meV.}
\label{fig:S3}
\end{figure}

Note that the decrease of the intensity centered at 2.6~meV is not an extrinsic effect.
This is confirmed by the data measured at different $E_i$ values. 
Extended Figure~\ref{fig:S3} shows a color contour map measured at 2.5~K with an $E_i$ of 3.14 meV.
The same dispersion relations as those measured with an $E_i$ of 5.9~meV are obtained,
indicating that the gap between two triplons bands is intrinsic.

\begin{table}[htb]
\caption{Crystal data for Ba$_2$CuSi$_2$O$_6$Cl$_2$.}
\label{table:1}
\begin{ruledtabular}
\begin{tabular}{cccc}
& Chemical formula & Ba$_2$CuSi$_2$O$_6$Cl$_2$ &  \\
& Space group & $Cmc2_1$ &  \\
& $a$ ($\rm{\AA}$) & 13.9064(3) &  \\
& $b$ ($\rm{\AA}$) & 13.8566(3) &  \\
& $c$ ($\rm{\AA}$) & 19.5767(4) &  \\
& $V$ ($\rm{\AA}^3$) & 3772.34(14) &  \\
& $Z$ & 16 &  \\
& $R;\ wR$ &  0.0376;\ 0.0803 & 
\end{tabular}
\end{ruledtabular}
\end{table}

\begin{table}[t]
\caption{Fractional atomic coordinates (${\times}\,10^4$) and equivalent isotropic displacement parameters ($\rm{\AA}^2{\times}\,10^3$) for Ba$_2$CuSi$_2$O$_6$Cl$_2$.}
\label{table:2}
\begin{ruledtabular}
\begin{tabular}{rrrrr}
Atom  &  $x$\hspace{7mm}    & $y$\hspace{7mm}   & $z$\hspace{5mm}  & $U_{\rm eq}$\hspace{1mm}   \\ \hline
Ba(1) & 5000 & 6058(1) & 3535(1) & 16(1)\\
Ba(2) & 5000 & 1445(1) & 3547(1) & 16(1)\\
Ba(3) & 7283(1) & 3787(1) & 6398(1) & 14(1)\\
Ba(4) & 7715(1) & 3684(1) & 3587(1) & 15(1)\\
Ba(5) & 5000 & 1005(1) & 6436(1) & 16(1)\\
Ba(6) & 5000 & 6383(1) & 6459(1) & 15(1)\\
Cu(1) & 7496(1) & 6219(1) & 4252(2) & 13(1)\\
Cu(2) & 7494(1) & 1250(1) & 5736(1) & 6(1)\\
Si(1) & 6109(2) & 2610(2) & 5010(2) & 9(1)\\
Si(2) & 6112(2) & 4864(2) & 4973(2) & 8(1)\\
Si(3) & 8889(2) & 4854(2) & 4978(2) & 8(1)\\
Si(4) & 8891(2) & 2600(2) & 5021(2) & 8(1)\\
O(1) & 5000 & 2433(6) & 4761(5) & 12(2)\\
O(2) & 6747(5) & 2389(5) & 4343(4) & 13(1)\\
O(3) & 6311(4) & 2025(4) & 5703(4) & 11(1)\\
O(4) & 6241(5) & 3746(3) & 5219(4) & 13(2)\\
O(5) & 6746(5) & 5045(5) & 4296(4) & 13(1)\\
O(6) & 5000 & 5029(6) & 4731(5) & 11(2)\\
O(7) & 6323(5) & 5495(4) & 5650(4) & 11(1)\\
O(8) & 8678(5) & 5448(5) & 4285(4) & 16(2)\\
O(9) & 10000 & 5035(7) & 5217(5) & 15(2)\\
O(10) & 8264(5) & 5073(5) & 5649(4) & 13(1)\\
O(11) & 8764(5) & 3724(4) & 4767(4) & 15(2)\\
O(12) & 10000 & 2445(6) & 5273(5) & 14(2)\\
O(13) & 8692(5) & 1975(4) & 4346(4) & 14(1)\\
O(14) & 8252(5) & 2433(5) & 5694(4) & 14(1)\\
Cl(1) & 7460(4) & 6440(4) & 2943(3) & 50(1)\\
Cl(2) & 5000 & 6015(4) & 1893(3) & 35(1)\\
Cl(3) & 5000 & 1260(3) & 1908(5) & 65(3)\\
Cl(4) & 5000 & 1370(5) & 8056(4) & 51(2)\\
Cl(5) & 7493(3) & 1052(4) & 7054(3) & 48(1)\\
Cl(6) & 5000 & 6294(3) & 8089(4) & 59(2)
\end{tabular}
\end{ruledtabular}
\end{table}

{\bf Derivation of dispersion relations}
In this section, we start from the model described in Fig.~\ref{model}a and derive the dispersion relations of triplet excitations.
The spin Hamiltonian is given by
\begin{equation}
\begin{split}
{\cal H} &= {\cal H}_0 + {\cal H}^{\prime}, \\
{\cal H}_0 &= J \sum_m \sum_n \boldsymbol{S}_{m n 1} \cdot \boldsymbol{S}_{m n 2}, \\
{\cal H}^{\prime} &= \sum_{m+n = \mathrm{even}} \sum_{\langle i,j\rangle} \Big( J^a_{ij} \boldsymbol{S}_{mni} \cdot \boldsymbol{S}_{(m+1)nj} \\
& \ \ \ +  J^{a\prime}_{ij} \boldsymbol{S}_{(m-1)n i} \cdot \boldsymbol{S}_{mnj} \\
& \ \ \ + J^b_{ij} \boldsymbol{S}_{mni} \cdot \boldsymbol{S}_{m(n+1)j} + J^b_{ij} \boldsymbol{S}_{m(n-1)i} \cdot \boldsymbol{S}_{mnj} \Big),
\label{Hamiltonian}
\end{split}
\end{equation}
where ${\cal H}_0$ represents intradimer exchange terms from $J$ and
${\cal H}^\prime$ represents interdimer exchange terms from $J^\xi_{ij}$ and $J^{\xi\prime}_{ij}$ ($\xi =~a, b$). 
$\boldsymbol{S}_{mni}$ is defined as the $i$-th Cu atom of the ($m$, $n$)-th dimer pair (see Fig.~\ref{model}a in the main text).
Dimers on the two different sublattices are distinguished by $m$ and $n$: $m+n$ becomes even for one sublattice and odd for the other. 
For each dimer pair $(m, n)$, a spin operator ${\bm S}_{mn}$, ${\bm T}_{mn}$ can be defined as
%For each dimer pair $(m, n)$, we introduce spin operators ${\bm S}_{mn}^\mathrm{e}$, ${\bm S}_{mn}^\mathrm{o}$, ${\bm T}_{mn}^\mathrm{e}$, ${\bm T}_{mn}^\mathrm{o}$ defined as
\begin{equation}
{\bm S}_{mn} \equiv {\bm S}_{mn1}+{\bm S}_{mn2},\hspace{2mm} {\bm T}_{mn} \equiv {\bm S}_{mn1}-{\bm S}_{mn2}.
%{\bm S}_{mn}^\mathrm{e}={\bm S}_{mn1}+{\bm S}_{mn2},\hspace{2mm} {\bm T}_{mn}^\mathrm{e}={\bm S}_{mn1}-{\bm S}_{mn2},
\end{equation}
%if $m+n =$~even, and
%\begin{equation}
%{\bm S}_{mn}^\mathrm{o}={\bm S}_{mn1}+{\bm S}_{mn2},\hspace{2mm} {\bm T}_{mn}^\mathrm{o}={\bm S}_{mn1}-{\bm S}_{mn2},
%\end{equation}
%if $m+n =$~odd.
Thus, ${\cal H}_0$ is rewritten as
\begin{equation}
{\cal H}_0 = \frac{J}{2} \sum_{m+n = \mathrm{even}} \Big( \boldsymbol{S}^2_{mn} + \boldsymbol{S}^2_{(m+1)n} - 3 \Big),
%{\cal H}_0 =~\sum_{m+n = \mathrm{even}} \frac{J}{2} \Big( \boldsymbol{S}^{\mathrm{e} \ 2}_{mn} + \boldsymbol{S}^{\mathrm{o} \ 2}_{mn} - 3 \Big), 
\label{H0dimer}
\end{equation}
since ${\bm S}^2_{mn1} = {\bm S}^2_{mn2} = 3/4$. In addition, ${\cal H}^\prime$ can be projected in a subspace constructed by the basis of ${\cal H}_0$ as
\begin{equation}
\begin{split}
{\cal H}^{\prime} &= {\cal H}^{\prime}_\mathrm{tt} + {\cal H}^{\prime}_\mathrm{ss}, \\
{\cal H}^{\prime}_\mathrm{tt} &=~\sum_{m+n = \mathrm{even}} \Big( J^A \boldsymbol{T}_{mn} \cdot \boldsymbol{T}_{(m+1)n} \\
& \ \ \ + J^{A\prime} \boldsymbol{T}_{(m-1)n} \cdot \boldsymbol{T}_{mn} \\
& \ \ \ + J^{B} \boldsymbol{T}_{mn} \cdot \boldsymbol{T}_{m(n+1)} + J^{B} \boldsymbol{T}_{m(n-1)} \cdot \boldsymbol{T}_{mn} \Big), \\
{\cal H}^{\prime}_\mathrm{ss} &=~\sum_{m+n = \mathrm{even}} \Big( J^A_\mathrm{s} \boldsymbol{S}_{mn} \cdot \boldsymbol{S}_{(m+1)n} \\
& \ \ \ + J^{A\prime}_\mathrm{s} \boldsymbol{S}_{(m-1)n} \cdot \boldsymbol{S}_{mn} \\
& \ \ \ + J^{B}_\mathrm{s} \boldsymbol{S}_{mn} \cdot \boldsymbol{S}_{m(n+1)} + J^{B}_\mathrm{s} \boldsymbol{S}_{m(n-1)} \cdot \boldsymbol{S}_{mn} \Big),
\label{Hpdimer}
\end{split}
\end{equation}
where 
\begin{equation}
\begin{split}
J^A &= \frac{1}{4}(J^a_{11} - J^a_{12} - J^a_{21} + J^a_{22}), \\
J^{A\prime} &= \frac{1}{4}(J^{a\prime}_{11} - J^{a\prime}_{12} - J^{a\prime}_{21} + J^{a\prime}_{22}), \\
J^B &= \frac{1}{4}(J^b_{11} - J^b_{12} - J^b_{21} + J^b_{22}), \\
J^A_\mathrm{s} &= \frac{1}{4}(J^a_{11} + J^a_{12} + J^a_{21} + J^a_{22}), \\
J^{A\prime}_\mathrm{s} &= \frac{1}{4}(J^{a\prime}_{11} + J^{a\prime}_{12} + J^{a\prime}_{21} + J^{a\prime}_{22}), \\
J^B_\mathrm{s} &= \frac{1}{4}(J^b_{11} + J^b_{12} + J^b_{21} + J^b_{22}).
\label{definitionofJ}
\end{split}
\end{equation}

Dispersion relations are obtained by applying a bond-operator approach \cite{bd, Matsumoto1, Matsumoto2} to eqs.~\eqref{H0dimer} and \eqref{Hpdimer}.
Singlet and triplet creation operators are defined as
\begin{equation}
\begin{split}
s^\dagger_{mn} |0 \rangle &=~\frac{1}{\sqrt{2}} (|\uparrow \rangle_{mn1} |\downarrow \rangle_{mn2} - |\downarrow \rangle_{mn1} |\uparrow \rangle_{mn2}), \\
t^\dagger_{xmn} |0 \rangle &=~-\frac{1}{\sqrt{2}}~(|\uparrow \rangle_{mn1} |\uparrow \rangle_{mn2} \\
& \ \ \ \ \ \ \ \ \ \ \ \ \ \ \ \ \ \ \ \ - |\downarrow \rangle_{mn1} |\downarrow \rangle_{mn2}), \\
t^\dagger_{ymn} |0 \rangle &=~\frac{i}{\sqrt{2}}~(|\uparrow \rangle_{mn1} |\uparrow \rangle_{mn2} + |\downarrow \rangle_{mn1} |\downarrow \rangle_{mn2}), \\
t^\dagger_{zmn} |0 \rangle &=~\frac{1}{\sqrt{2}} (|\uparrow \rangle_{mn1} |\downarrow \rangle_{mn2} + |\downarrow \rangle_{mn1} |\uparrow \rangle_{mn2}),
\end{split}
\end{equation}
so that they follow bosonic commutation relations. In this definition, the number of bosons per dimer is constrained to 1 as
\begin{equation}
s^\dagger_{mn} s_{mn} + \sum_{\alpha=x,y,z} t^\dagger_{\alpha mn} t_{\alpha mn} = 1.
\end{equation}
Then, the squared operator $\boldsymbol{S}^2_{mn}$ and $\alpha$ component ($\alpha =~x,y,z$) of $\boldsymbol{S}_{mn}$ and $\boldsymbol{T}_{mn}$ are given as
\begin{equation}
\begin{split}
\boldsymbol{S}^2_{mn} &= \sum_{\alpha=x,y,z} t^\dagger_{\alpha mn} t_{\alpha mn}, \\
S_{\alpha mn} &= -i \sum_{\beta,\gamma=x,y,z} \epsilon_{\alpha \beta \gamma} t^\dagger_{\beta mn} t_{\gamma mn}, \\  
T_{\alpha mn} &= s^\dagger_{mn} t_{\alpha mn} +  t^\dagger_{\alpha mn} s_{mn},
%\boldsymbol{S}^2_{mn} &=2 (t^\dagger_{+,mn} t_{+,mn} + t^\dagger_{0,mn} t_{0,mn} + t^\dagger_{-,mn} t_{-,mn}) \\
%\boldsymbol{S}_{mn} \cdot \boldsymbol{S}_{m^\prime n^\prime} &\rightarrow S_{mn,m^\prime n^\prime} \\
%S_{mn,m^\prime n^\prime} &\equiv t^\dagger_{0,mn} t_{0,m^\prime n^\prime} (t_{+,mn} t^\dagger_{+,m^\prime n^\prime} + t_{-,mn} t^\dagger_{-,m^\prime n^\prime}) + (h.c.) \\
%& \ \ \ - t^\dagger_{0,mn} t^\dagger_{0,m^\prime n^\prime} (t_{+,mn} t_{-,m^\prime n^\prime} + t_{-,mn} t_{+,m^\prime n^\prime}) + (h.c.) \\
%& \ \ \ +  (t^\dagger_{+,mn} t_{+,mn} - t^\dagger_{-,mn} t_{-,mn}) (t^\dagger_{+,m^\prime n^\prime} t_{+,m^\prime n^\prime} - t^\dagger_{-,m^\prime n^\prime} t_{-,m^\prime n^\prime}) \\
%\boldsymbol{T}_{mn} \cdot \boldsymbol{T}_{m^\prime n^\prime} &\rightarrow T_{mn,m^\prime n^\prime} \\
%T_{mn,m^\prime n^\prime} &\equiv (t^\dagger_{+,mn} t_{+,m^\prime n^\prime} + t^\dagger_{0,mn} t_{0,m^\prime n^\prime} + t^\dagger_{-,mn} t_{-,m^\prime n^\prime}) s^\dagger_{mn} s_{m^\prime n^\prime} + (h.c.) \\
%& \ \ \ +  (t^\dagger_{+,mn} t^\dagger_{-,m^\prime n^\prime} + t^\dagger_{0,mn} t^\dagger_{0,m^\prime n^\prime} + t^\dagger_{-,mn} t^\dagger_{+,m^\prime n^\prime}) s_{mn} s_{m^\prime n^\prime} + (h.c.)
\end{split}
\end{equation}
where $\epsilon_{\alpha \beta \gamma}$ represents an antisymmetric tensor.
At zero field, the ground state is a product of the singlet at each dimer, and thus, the triplon density is zero.
Thus, a mean-field approximation that neglects the dynamics of singlet operators should be applicable. 
By replacing creation and annihilation operators by its expectation value, $\langle s^\dagger_{mn} \rangle \sim \langle s_{mn} \rangle \sim 1$, and neglecting high-order terms,
eqs.~\eqref{H0dimer} and \eqref{Hpdimer} become
\begin{align}
%{\cal H}_0 &= \sum_{m+n = \mathrm{even}} \sum_{\alpha=+,0,-} J \Big( t^\dagger_{\alpha mn} t_{\alpha,m^\prime n^\prime} - \frac{3}{4} \Big) \\
{\cal H}_0 &= J \sum_{m+n = \mathrm{even}} \Big( \sum_{\alpha=x,y,z} t^\dagger_{\alpha mn} t_{\alpha mn} - \frac{3}{4} \Big), \notag \\
%{\cal H}^{\prime}_\mathrm{tt} &= \sum_{m+n = \mathrm{even}} (J^A T_{mn,(m+1)n} + J^{A\prime} T_{(m-1)n,mn} + J^B T_{mn,m(n+1)} + J^B T_{m(n-1),mn})
{\cal H}^{\prime} &\sim \sum_{m+n = \mathrm{even}} \sum_{\alpha=x,y,z} \Big[ J^A \big\{ t^\dagger_{\alpha mn} t_{\alpha (m+1)n} \notag \\
&+ t^\dagger_{\alpha mn} t^\dagger_{\alpha (m+1)n} + (h.c.) \big\} \label{realH} \\
&+ J^{A\prime} \big\{ t^\dagger_{\alpha (m-1)n} t_{\alpha mn} + t^\dagger_{\alpha (m-1)n} t^\dagger_{\alpha mn} + (h.c.) \big\} \notag \\
&+ J^B \big\{ t^\dagger_{\alpha mn} t_{\alpha m(n+1)} + t^\dagger_{\alpha mn} t^\dagger_{\alpha m(n+1)} + (h.c.) \big\} \notag \\
&+ J^B \big\{ t^\dagger_{\alpha m(n-1)} t_{\alpha mn} + t^\dagger_{\alpha m(n-1)} t^\dagger_{\alpha mn} + (h.c.) \big\} \Big]. \notag
%{\cal H}^{\prime}_\mathrm{tt} &= \sum_{m+n = \mathrm{even}} J^A (t^\dagger_{+,mn} t_{+,(m+1)n} + t^\dagger_{0,mn} t_{0,(m+1)n} \\
%& \ \ \ + t^\dagger_{-,mn} t_{-,(m+1)n} + t^\dagger_{+,mn} t^\dagger_{-,(m+1)n}  \\
%& \ \ \ + t^\dagger_{0,mn} t^\dagger_{0,(m+1)n} + t^\dagger_{-,mn} t^\dagger_{+,(m+1)n}) + (h.c.) \\
%& \ \ \ + J^{A\prime} (t^\dagger_{+,(m-1)n} t_{mn} + t^\dagger_{0,(m-1)n} t_{0,mn} \\
%& \ \ \ + t^\dagger_{-,(m-1)n} t_{-,mn} + t^\dagger_{+,(m-1)nn} t^\dagger_{-,mn} \\
%& \ \ \ + t^\dagger_{0,(m-1)n} t^\dagger_{0,mn} + t^\dagger_{-,(m-1)n} t^\dagger_{+,mn}) + (h.c.) \\
%& \ \ \ + J^B (t^\dagger_{+,mn} t_{+,m(n+1)} + t^\dagger_{0,mn} t_{0,,m(n+1)} \\
%& \ \ \ + t^\dagger_{-,mn} t_{-,,m(n+1)} + t^\dagger_{+,mn} t^\dagger_{-,,m(n+1)} \\ 
%& \ \ \ + t^\dagger_{0,mn} t^\dagger_{0,,m(n+1)} + t^\dagger_{-,mn} t^\dagger_{+,,m(n+1)}) + (h.c.) \\
%& \ \ \ + J^{B\prime} (t^\dagger_{+,m(n-1)} t_{mn} + t^\dagger_{0,m(n-1)} t_{0,mn} \\
%& \ \ \ + t^\dagger_{-,m(n-1)} t_{-,mn} + t^\dagger_{+,m(n-1)} t^\dagger_{-,mn} \\
%& \ \ \ + t^\dagger_{0,m(n-1)} t^\dagger_{0,mn} + t^\dagger_{-m(n-1)} t^\dagger_{+,mn}) + (h.c.)
\end{align}

A $\mathbf{k}$-dependent form is obtained by Fourier transformation defined at each sublattice as
\begin{equation}
\begin{split}
t^{\dagger}_{\alpha mn} = \sqrt{\frac{2}{N}} \sum_\mathbf{k} e^{i \mathbf{k} \cdot \mathbf{r}_{mn}} t^{\dagger, 1}_{\alpha\mathbf{k}}, \\
t_{\alpha mn} = \sqrt{\frac{2}{N}} \sum_\mathbf{k} e^{-i \mathbf{k} \cdot \mathbf{r}_{mn}} t^{1}_{\alpha\mathbf{k}},
\end{split}
\end{equation}
for $m+n =$~even and 
\begin{equation}
\begin{split}
t^{\dagger}_{\alpha mn} &= \sqrt{\frac{2}{N}} \sum_\mathbf{k} e^{i \mathbf{k} \cdot \mathbf{r}_{mn}} t^{\dagger, 2}_{\alpha\mathbf{k}}, \\
t_{\alpha mn} &= \sqrt{\frac{2}{N}} \sum_\mathbf{k} e^{-i \mathbf{k} \cdot \mathbf{r}_{mn}} t^{2}_{\alpha\mathbf{k}},
\end{split}
\end{equation}
for $m+n =$~odd, where $N$ describes the number of dimers.
This procedure leads to the following quadratic form:
\begin{equation}
\begin{split}
{\cal H}_0 &= J \sum_{\mathbf{k}} \Big( \sum_{\alpha=x,y,z} \ t^\dagger_{\alpha\mathbf{k}} t_{\alpha\mathbf{k}} - \frac{3}{4} \Big), \\
{\cal H}^{\prime} &= \sum_{\mathbf{k}} \sum_{\alpha=x,y,z} \Lambda_\mathbf{k} \big\{ t^{\dagger,1}_{\alpha\mathbf{k}} t^{2}_{\alpha\mathbf{k}} + t^{\dagger,1}_{\alpha\mathbf{k}} t^{\dagger, 2}_{\alpha\mathbf{k}} + (h.c.) \big\},
\label{last}
\end{split}
\end{equation}
where 
\begin{equation}
\begin{split}
\Lambda_\mathbf{k} &= J^A e^{-i k_x a/2} + J^{A\prime} e^{i k_x a/2} \\ 
&\ \ \ \ \ + J^B \big( e^{-i k_y b/2} + e^{i k_y b/2} \big).
\end{split}
\end{equation}

Eq.~\eqref{last} can be described using a 4 $\times$ 4 matrix as
\begin{equation}
\begin{split}
{\cal H} &= {\cal H}_0 + {\cal H}^\prime = \frac{1}{2} \sum_\mathbf{k} \sum_{\alpha=x,y,z} {\cal H_\alpha} - \frac{3}{4} NJ, \\
{\cal H_\alpha} &= \left( t^{\dagger, 1}_{\alpha\mathbf{k}} t^{\dagger, 2}_{\alpha(-\mathbf{k})} t^1_{\alpha\mathbf{k}} t^2_{\alpha(-\mathbf{k})} \right)
{\cal M}_\mathbf{k}
\left( \begin{array}{c} t^1_{\alpha\mathbf{k}} \\ t^2_{\alpha\mathbf{k}} \\ t^{\dagger, 1}_{\alpha(-\mathbf{k})} \\ t^{\dagger, 2}_{\alpha(-\mathbf{k)}} \end{array} \right),
\end{split}
\end{equation}
which is the same as eq.~(1) in the main text (except for the omitted constant term), where
\begin{equation}
\begin{split}
{\cal M}_\mathbf{k} =~\left( ~\begin{matrix} J & \Lambda_\mathbf{k} & 0 & \Lambda_\mathbf{k} \\ \Lambda^*_\mathbf{k} & J & \Lambda^*_\mathbf{k} & 0 \\
0 & \Lambda_\mathbf{k} & J & \Lambda_\mathbf{k} \\ \Lambda^*_\mathbf{k} & 0 & \Lambda^*_\mathbf{k} & J \end{matrix} \right).
\end{split}
\end{equation}
\\
The dispersion relation can be obtained by Bogoliubov transformation, which is equivalent to a procedure determining a paraunitary matrix $T_\mathbf{k}$ that satisfies
%$T_\mathbf{k}^\dagger {\cal M}_\mathbf{k} T_\mathbf{k} =~\mathrm{diag}(E_{+, \mathbf{k}}, E_{-, \mathbf{k}}, E_{+, -\mathbf{k}}, E_{-, -\mathbf{k}})$,
\begin{equation}
\begin{split}
T_\mathbf{k}^\dagger {\cal M}_\mathbf{k} T_\mathbf{k} = \left( \begin{matrix} E_{+, \mathbf{k}} & 0 & 0 & 0 \\ 0 & E_{-, \mathbf{k}} & 0 & 0 \\
0 & 0 &  E_{+, -\mathbf{k}} & 0 \\ 0 & 0 & 0 & E_{-, -\mathbf{k}} \end{matrix} \right).
\label{diagonalized}
\end{split}
\end{equation}
Owing to orthogonality and completeness of the new basis, $T_\mathbf{k}^\dagger \boldsymbol{\Sigma} T_\mathbf{k} = T_\mathbf{k} \boldsymbol{\Sigma} T_\mathbf{k}^\dagger = ~\boldsymbol{\Sigma}$,
where $\boldsymbol{\Sigma} \equiv~\mathrm{diag}(1,1,-1,-1)$. Therefore, eq.~\eqref{diagonalized} is equivalent to the relation
%$T_\mathbf{k}^\dagger {\cal M}_\mathbf{k} T_\mathbf{k} =~\mathrm{diag}(E_{+, \mathbf{k}}, E_{-, \mathbf{k}}, E_{+, -\mathbf{k}}, E_{-, -\mathbf{k}})$,
\begin{equation}
\begin{split}
\boldsymbol{\Sigma} {\cal M}_\mathbf{k} T_\mathbf{k} = T_\mathbf{k} \boldsymbol{\Sigma} \left( ~\begin{matrix} E_{+, \mathbf{k}} & 0 & 0 & 0 \\ 0 & E_{-, \mathbf{k}} & 0 & 0 \\
0 & 0 & E_{+, -\mathbf{k}} & 0 \\ 0 & 0 & 0 & E_{-, -\mathbf{k}} \end{matrix} \right).
\end{split}
\end{equation}
Thus, eigenenergies $E_{+, \mathbf{k}}, E_{-, \mathbf{k}}, -E_{+, \mathbf{k}}$, and $-E_{-, \mathbf{k}}$ are obtained by diagonalizing $\boldsymbol{\Sigma} {\cal M}_\mathbf{k}$, leading to the dispersion relation given by
eq.~\eqref{diseq} in the main text.
%\begin{equation}
%E_{\pm, \mathbf{k}} =~\sqrt{J^2 \pm 2 J |\Lambda_\mathbf{k}|}. \label{disp}
%\end{equation}

{\bf Calculation of Berry connection}
In this section, we start by determining $T_\mathbf{k}$ and then derive the Berry connection of each subband from the Hamiltonian ${\cal M}^{(4)}_\mathbf{k}$ (eq.~\eqref{gH}).
By diagonalizing $\boldsymbol{\Sigma} {\cal M}^{(4)}_\mathbf{k}$, eigenvectors for each eigenenergy are determined as
\begin{equation}
\begin{split}
E &=~E_{+, \mathbf{k}} \equiv \sqrt{J^2 + 2 J |\boldsymbol{d}|}: \\
&\mathbf{t}_{++, \mathbf{k}} =~\frac{1}{\Delta^+_\mathbf{k}} \left(\begin{matrix}
A_\mathbf{k} (|\boldsymbol{d}| + d_z) \\ A_\mathbf{k} (d_x + i d_y) \\ B_\mathbf{k} (|\boldsymbol{d}| + d_z) \\ B_\mathbf{k} (d_x + i d_y) \end{matrix}\right), \\
E &=~E_{-, \mathbf{k}} \equiv \sqrt{J^2 - 2 J |\boldsymbol{d}|}: \\
& \mathbf{t}_{+-, \mathbf{k}} =~\frac{1}{\Delta^-_\mathbf{k}} \left(\begin{matrix}
C_\mathbf{k} (-d_x + i d_y) \\ C_\mathbf{k} (|\boldsymbol{d}| + d_z) \\ D_\mathbf{k} (-d_x + i d_y) \\ D_\mathbf{k} (|\boldsymbol{d}| + d_z) \end{matrix}\right), \\
E &=~-E_{-, -\mathbf{k}} = -\sqrt{J^2 - 2 J |\boldsymbol{d}|}: \\
& \mathbf{t}_{-+, -\mathbf{k}} =~\frac{1}{\Delta^-_\mathbf{k}} \left(\begin{matrix}
D_\mathbf{k} (-d_x + i d_y) \\ D_\mathbf{k} (|\boldsymbol{d}| + d_z) \\ C_\mathbf{k} (-d_x + i d_y) \\ C_\mathbf{k}  (|\boldsymbol{d}| + d_z) \end{matrix}\right), \\
E &=~-E_{+, -\mathbf{k}} = -\sqrt{J^2 + 2 J |\boldsymbol{d}|}: \\
& \mathbf{t}_{--, -\mathbf{k}} =~\frac{1}{\Delta^+_\mathbf{k}} \left(\begin{matrix}
B_\mathbf{k} (|\boldsymbol{d}| + d_z) \\ B_\mathbf{k} (d_x + i d_y) \\ A_\mathbf{k}  (|\boldsymbol{d}| + d_z) \\ A_\mathbf{k} (d_x + i d_y) \end{matrix}\right),
\label{eigenvectors}
\end{split}
\end{equation}
where
\begin{equation}
\begin{split}
A_\mathbf{k} &=~J + \sqrt{J^2 + 2 J |\boldsymbol{d}|}, \\
B_\mathbf{k} &=~J - \sqrt{J^2 + 2 J |\boldsymbol{d}|}, \\
C_\mathbf{k} &=~J + \sqrt{J^2 - 2 J |\boldsymbol{d}|}, \\
D_\mathbf{k} &=~J - \sqrt{J^2 - 2 J |\boldsymbol{d}|}, \\
\Delta^{+2}_\mathbf{k} &=~2 (A_\mathbf{k}^2 - B_\mathbf{k}^2) |\boldsymbol{d}| (|\boldsymbol{d}| + d_z) \\
&=~8 |\boldsymbol{d}| (|\boldsymbol{d}| + d_z) J \sqrt{J^2 + 2 J |\boldsymbol{d}|}, \\
\Delta^{-2}_\mathbf{k} &=~2 (C_\mathbf{k}^2 - D_\mathbf{k}^2) |\boldsymbol{d}| (|\boldsymbol{d}| + d_z) \\
&=~8 |\boldsymbol{d}| (|\boldsymbol{d}| + d_z) J \sqrt{J^2 - 2 J |\boldsymbol{d}|}.
\end{split}
\end{equation}
Thus, a paraunitary matrix can be constructed as $T_\mathbf{k} =~(\mathbf{t}_{++, \mathbf{k}}, \mathbf{t}_{+-, \mathbf{k}}, \mathbf{t}_{-+, -\mathbf{k}}, \mathbf{t}_{--, -\mathbf{k}})$.
Note that this definition is not valid and a different gauge should be selected for $\boldsymbol{d} = (0, 0, -d) (d > 0)$ .
The following discussion can be also applied to eigenvectors with a different gauge.

The Berry connection can be defined by the following equation~\cite{magnon1, MagnonHall3},
\begin{equation}
A_{j\mu, \mathbf{k}} =~ -i \mathrm{Tr} \left[ \Gamma_j \Sigma T_\mathbf{k}^\dagger \Sigma \frac{\partial T_\mathbf{k}}{\partial k_\mu} \right],
\label{Berrycurvature}
\end{equation}
where $\Gamma_j$ is a diagonal matrix, the $j$-th diagonal component of which is 1 while others are zero, and $\mu = x, y$.
From eq.~\eqref{Berrycurvature}, the Berry connection of each subband for $\mu = x$ can also be rewritten as
\begin{equation}
\begin{split}
A_{++, \mathbf{k}} =~ -i \mathbf{t}^\dagger_{++, \mathbf{k}} \Sigma \frac{\partial \mathbf{t}_{++, \mathbf{k}}}{\partial k_x}, \\
A_{+-, \mathbf{k}} =~ -i \mathbf{t}^\dagger_{+-, \mathbf{k}} \Sigma \frac{\partial \mathbf{t}_{+-, \mathbf{k}}}{\partial k_x}, \\
A_{-+, \mathbf{k}} =~ i \mathbf{t}^\dagger_{-+, -\mathbf{k}} \Sigma \frac{\partial \mathbf{t}_{-+, -\mathbf{k}}}{\partial k_x}, \\
A_{--, \mathbf{k}} =~ i \mathbf{t}^\dagger_{--, -\mathbf{k}} \Sigma \frac{\partial \mathbf{t}_{--, -\mathbf{k}}}{\partial k_x}.
\label{Amatirx}
\end{split}
\end{equation}
Substituting eqs.~\eqref{eigenvectors} for eqs.~\eqref{Amatirx} leads to
\begin{align}
A_{++, \mathbf{k}} &=~\frac{1}{2|\boldsymbol{d}|(|\boldsymbol{d}| + d_z)} \left( d_x \frac{\partial d_y}{\partial k_x} - d_y \frac{\partial d_x}{\partial k_x} \right) + \cdots, \notag \\
A_{+-, \mathbf{k}} &=~-\frac{1}{2|\boldsymbol{d}|(|\boldsymbol{d}| + d_z)} \left( d_x \frac{\partial d_y}{\partial k_x} - d_y \frac{\partial d_x}{\partial k_x} \right) + \cdots, \notag \\
A_{-+, \mathbf{k}} &= A_{+-, \mathbf{k}}, A_{--, \mathbf{k}} = A_{++, \mathbf{k}}.
\label{ReA}
\end{align}
The first real term corresponds to the phase change of the eigenvector along the Brillouin zone,
while the remaining of imaginary terms omitted in eq.~\eqref{ReA} are due to band deformation.
For a one-dimensional system, the total phase change across the Brillouin zone corresponds to the Zak phase \cite{Zak}:
\begin{equation}
\gamma_{j} = -\int_{\mathrm{BZ}} A_{j, \mathbf{k}} = -\int_{\mathrm{BZ}} \mathrm{Re} A_{j, \mathbf{k}}.
\end{equation}
Under $d_z = 0$, $\boldsymbol{d}$ can be represented by $(|\boldsymbol{d}| \cos {\theta}, -|\boldsymbol{d}| \sin {\theta}, 0)$, leading to
\begin{equation}
\begin{split}
\mathrm{Re}(A_{++, \mathbf{k}}) &=~-\frac{1}{2} \frac{\partial \theta}{\partial k_x}, \\
\gamma_{++} &=~-\int_{\mathrm{BZ}} A_{++, \mathbf{k}} = n \pi, \\
\gamma_{++} &= -\gamma_{+-} = -\gamma_{-+} = \gamma_{--},
\end{split}
\end{equation}
where the integer $n$ represents the winding number.
%\begin{equation}
%\mathrm{Re}(A_{++, \mathbf{k}}) =~-\frac{i}{2} \frac{\Lambda_\mathbf{k}}{|\Lambda_\mathbf{k}|} \frac{\partial}{\partial k_x} \left( \frac{\Lambda_\mathbf{k}^*}{|\Lambda_\mathbf{k}|} \right) = -\frac{1}{2} \frac{\partial \theta_\mathbf{k}}{\partial k_x}.
%\end{equation}
The exactly same form can be derived from ${\cal M}_\mathbf{k}^{(2)} = J\mathbf{1} + \boldsymbol{d} \cdot \boldsymbol{\sigma}$ for an arbitrary gauge,
indicating that topological properties are unchanged even if pair creation and annihilation terms are present.

For triplon bands in Ba$_2$CuSi$_2$O$_6$Cl$_2$, the Berry connection can be obtained from $\boldsymbol{d} =$ (Re$\Lambda_\mathbf{k}$, $-$Im$\Lambda_\mathbf{k}$, 0) as
\begin{align}
A_{++, \mathbf{k}} &=~-\frac{i}{2} \frac{\Lambda_\mathbf{k}}{|\Lambda_\mathbf{k}|} \frac{\partial}{\partial k_x} \left( \frac{\Lambda_\mathbf{k}^*}{|\Lambda_\mathbf{k}|} \right) \notag \\
&\ + i \Bigg\{ \frac{2B_\mathbf{k}}{\Delta^+_\mathbf{k}} \frac{\partial}{\partial k_x}\left(\frac{B_\mathbf{k}}{\Delta^+_\mathbf{k}}\right) - \frac{2A_\mathbf{k}}{\Delta^+_\mathbf{k}} \frac{\partial}{\partial k_x} \left( \frac{A_\mathbf{k}}{\Delta^+_\mathbf{k}} \right)  \Bigg\}, \notag \\
A_{+-, \mathbf{k}} &=~-\frac{i}{2} \frac{\Lambda^*_\mathbf{k}}{|\Lambda_\mathbf{k}|} \frac{\partial}{\partial k_x} \left( \frac{\Lambda_\mathbf{k}}{|\Lambda_\mathbf{k}|} \right) \\
&\ + i \Bigg\{ \frac{2D_\mathbf{k}}{\Delta^-_\mathbf{k}} \frac{\partial}{\partial k_x}\left(\frac{D_\mathbf{k}}{\Delta^-_\mathbf{k}}\right) - \frac{2C_\mathbf{k}}{\Delta^-_\mathbf{k}} \frac{\partial}{\partial k_x} \left( \frac{C_\mathbf{k}}{\Delta^-_\mathbf{k}} \right)\Bigg\}, \notag \\
A_{--, \mathbf{k}} &= A_{++, \mathbf{k}}, \notag \\
A_{-+, \mathbf{k}} &= A_{+-, \mathbf{k}}, \notag 
\end{align}
which leads to the Zak phase quantized into $\gamma_{++} = -\gamma_{+-} = -\gamma_{-+} = \gamma_{--} = \pm \pi$ irrespective of $k_y$.

{\bf Calculation of an energy spectrum}
As discussed in the main text, edge states should appear at the end of the $a$-direction from an analogy with a coupled SSH model~\cite{coupledSSH}.
To confirm this, an energy spectrum of the present model is calculated by imposing open boundary conditions along the $a$-direction.
For simplicity, Fourier-transformed operators are defined under periodic boundary conditions along the $b$-direction as
\begin{equation}
\begin{split}
t^{\dagger}_{\alpha mn} &= \sqrt{\frac{2}{N_b}} \sum_{k_y} e^{i k_y y_{mn}} t^{\dagger}_{\alpha mk_y}, \\
t_{\alpha mn} &= \sqrt{\frac{2}{N_b}} \sum_{k_y} e^{-i k_y y_{mn}} t_{\alpha mk_y}, \label{FTkb}
\end{split}
\end{equation}
where $N_b$ is the number of dimers in a single chain.
Substituting eqs.~\eqref{FTkb} for eqs.~\eqref{realH} leads to
\begin{equation}
\begin{split}
{\cal H} = \frac{1}{2} \sum_{k_y} \sum_{\alpha = x, y, z} \mathbf{m}^\dagger_{\alpha k_y} \left( \begin{matrix} J \mathbf{1} + \mathbf{X}_{k_y} & \mathbf{X}_{k_y} \\ \mathbf{X}_{k_y} & J \mathbf{1} + \mathbf{X}_{k_y} \end{matrix} \right) \mathbf{m}_{\alpha k_y},
\end{split}
\end{equation}
where $\mathbf{m}_{\alpha k_y}$ represents a $4N_a$ ($N_a \equiv N/N_b$) component vector
\begin{equation}
\begin{split}
\mathbf{m}_{\alpha k_y} &\equiv (t^{1, \dagger}_{\alpha mk_y}, t^{1, \dagger}_{\alpha mk_y}, \cdots, t^{N_a, \dagger}_{\alpha mk_y}, t^{N_a, \dagger}_{\alpha mk_y}, \\
&\ \ \ \ \ \ t^{1}_{\alpha m-k_y}, t^{1}_{\alpha m-k_y}, \cdots, t^{N_a}_{\alpha m-k_y}, t^{N_a}_{\alpha m-k_y})^T,
\end{split}
\end{equation}
and $\mathbf{1}$ is an $N_a \times N_a$ identity matrix. $\mathbf{X}_{k_y}$ is a $2N_a \times 2N_a$ matrix defined as
\begin{equation}
\begin{split}
\mathbf{X}_{k_y} = \left( \begin{matrix} 0 & J_{k_y} & 0 & J^{A\prime} & 0 & 0 & \cdots & 0 & 0  \\ J_{k_y} & 0 & J^A & 0 & 0 & 0 &\cdots & 0 & 0 \\
0 &J^A & 0 & J_{k_y} & 0 & J^{A\prime} & \cdots & 0 & 0 \\ J^{A\prime} & 0 & J_{k_y} & 0 & J^A & 0 & \cdots & 0 & 0 & \\
0 & 0 & 0 & J^A & 0 & J_{k_y} & \cdots & 0 & 0 & \\ 0 & 0 & J^{A\prime} & 0 & J_{k_y} & 0 & \cdots & 0 & 0 \\
\vdots & \vdots & \vdots & \vdots & \vdots & \vdots & \ddots & \vdots & \vdots \\ 
0 & 0 & 0 & 0 & 0 & 0 & \cdots & 0 & J_{k_y} \\ 0 & 0 & 0 & 0 & 0 & 0 & \cdots & J_{k_y} & 0 \end{matrix} \right), \label{Mmatrix}
\end{split}
\end{equation}
where $J_{k_y} = 2J_B \cos(k_y b/2)$.
The energy spectrum is obtained by diagonalizing the $4N_a \times 4N_a$ matrix in eq.~\eqref{Mmatrix} for each $k_y$.
Figure~\ref{edge}b in the main text represents the energy spectrum with $N_a$ = 100, which clearly exhibits a twofold-degenerate edge state at the energy $J$.
The bulk excitation spectrum is also consistent with the dispersion relation given by eq.~\eqref{diseq}.
%\end{document}
\end{document}